\documentclass[printer]{aa}  

\usepackage{graphicx}
\usepackage{bm}
\usepackage{xcolor}
\usepackage{multirow,multicol}
\usepackage{soul}

\usepackage{txfonts}
\usepackage[breaklinks, colorlinks, citecolor=blue]{hyperref}
\def\drm{\mathrm{d}}

\def\fesc{{f_\mathrm{esc}}}
\def\Mturn{{M_\mathrm{turn}}}

\def\xloc{{x_\mathrm{loc}}}
\def\sigloc{{\sigma_\mathrm{loc}}}
\def\siglocbright{{\sigma_\mathrm{loc}^{21}}}
\def\siglocion{{\sigma_\mathrm{loc}^\mathrm{ion}}}
\def\sigwhole{{\sigma_\mathrm{whole}}}

\begin{document} 

   \title{Using the sample variance of 21cm maps as a tracer\\of the ionisation topology}
\author{
    A. Gorce \inst{1,2,3} \and A. Hutter \inst{4} \and J. R. Pritchard\inst{2}
}

\institute{
    Université Paris-Saclay, CNRS,  
    Institut d'Astrophysique Spatiale, 
    91405, Orsay, France
    \and 
	Department of Physics,
	Blackett Laboratory, 
	Imperial College,			
	London SW7 2AZ, U.K.
	\and
    Department of Physics and McGill Space Institute, McGill University, Montreal, QC, Canada H3A 2T8 
     \and Kapteyn Astronomical Institute, University of Groningen, PO Box 800, NL-9700 AV Groningen, The Netherlands\\
	\email{adelie.gorce@mail.mcgill.ca}
}
\date{Received  8 February 2021 / Accepted 28 June 2021}

% \abstract{}{}{}{}{} 
% 5 {} token are mandatory
 
  \abstract
   {Intensity mapping of the 21cm signal of neutral hydrogen will yield exciting insights into the Epoch of Reionisation and the nature of the first galaxies. However, the large amount of data that will be generated by the next generation of radio telescopes, such as the Square Kilometre Array (SKA), as well as the numerous observational obstacles to overcome, require analysis techniques tuned to extract the reionisation history and morphology. In this context, we introduce a one-point statistic, which we refer to as the local variance, $\sigloc$, that describes the distribution of the mean differential 21cm brightness temperatures measured in two-dimensional maps along the frequency direction of a light cone. The local variance takes advantage of what is usually considered an observational bias, the sample variance. We find the redshift-evolution of the local variance to not only probe the reionisation history of the observed patches of the sky, but also trace the ionisation morphology. This estimator provides a promising tool to constrain the midpoint of reionisation as well as gain insight into the ionising properties of early galaxies.}


\maketitle
%
%________________________________________________________________

\section{Introduction}
\label{sec:intro}

The Epoch of Reionisation (EoR) represents an essential time within the first billion years of the Universe, when the first light sources formed and gradually ionised the neutral hydrogen gas in the intergalactic medium (IGM). However, the exact properties of these first light sources remain uncertain. With the help of the Cosmic Microwave Background (CMB) small- and large-scale data \citep{planck_2016_reion}, observations of the luminosity functions of star-forming galaxies \citep{bouwens_2015}, and $\ion{H}{I}$ absorption troughs in the spectra of quasars \citep{McGreerMesinger_2015,banados_2018}, we have obtained constraints on the ionisation history, that is the time evolution of the ionisation fraction of the hydrogen gas in the IGM  \citep{robertson_2015,gorce_2018}. However, reionisation is patchy, with different regions of the sky being ionised at different times.

For this reason, the observation of the time evolution of the 21cm brightness temperature maps at redshifts $z\geq 5$, tracing the neutral hydrogen gas in the IGM and referred to as 21cm tomography is highly anticipated and should be achieved with the next generation of radio interferometers, such as the Square Kilometre Array \citep[SKA,][]{SKA} or the Hydrogen Epoch of Reionization Array \citep[HERA,][]{HERA}. Such maps will allow us to trace the reionisation process, in particular the time and spatial evolution of the ionised regions around the first light sources. Constraints from current observations imply that star-forming galaxies were the main drivers of reionisation. The topology of the ionised regions provides a tracer of the physical properties of these galaxies and their distribution in the IGM \citep{ZahnLidz_2007, McQuinnLidz_2007,21cmFAST_2011}. In the recent years, many statistical tools of various levels of complexity have been developed to extract information about the ionisation sources from these maps, but this is complicated by the cosmological signal often being swamped with thermal noise and foregrounds \citep{ChapmanJelic_2019,LiuShaw_2020,HothiChapman_2021,Gagnon-HartmanCui_2021}. 
While the power spectrum of the 21cm signal has been the main tool to extract the Gaussian part of the 21cm signal from reionisation \citep[e.g.][]{GreigTrott_2020,PaganoLiu_2020}, three-point statistics have been increasingly studied to access the non-Gaussian part of the signal \citep{ShimabukuroYoshiura_2016,MajumdarPritchard_2018,GorcePritchard_2019,watkinson_2019_bispectrum,hutter_2020_bispectrum}. %However, three-point statistics are often computationally expensive and challenging to interpret. One-point statistics provide here a faster approach.

Choosing a simpler approach, many works have focussed on the one-point probability distribution function (PDF) of the differential 21cm brightness temperature $\delta T_b$ and its moments \citep{CiardiMadau_2003,furlanetto_2004_excursion_set, MellemaIliev_2006}. Since the morphology of this 21cm signal is driven by the morphology of the ionised regions during the EoR, it is informative to assess the PDF of the ionisation fraction distribution. For example, for a binary ionisation field, where pixels are either fully ionised or fully neutral, the corresponding one-point PDF can be derived as a combination of Dirac delta functions $\delta$:
\begin{equation}
\label{eq:def_PDF_3D_pixels}
    P(x_e) = (1-\bar{x}_e)\, \delta (x_e) + \bar{x}_e\, \delta (x_e-1),
\end{equation}
where $\bar{x}_e$ is the filling fraction -- or the mean ionisation level of the simulation. From this PDF, analytic expression for statistical moments can be derived. Comparing how these statistical moments, derived numerically from more sophisticated models and simulations, deviate from these analytical expression
provide us with hints on the nature of reionisation \citep{Gluscevic_Barkana_2010}, such as its reionisation topology \citep[e.g. inside-out or outside-in, see][]{watkinson_2014_1PCF} and its global reionisation history \citep{BittnerLoeb_2011,Patil_2014}, even when derived from dirty 21cm signal images or after foreground removal \citep{HarkerZaroubi_2009}. \citet{KittiwisitBowman_2018} show that HERA will be able to detect the variance of the 21cm brightness temperature field from reionisation with high sensitivity, by averaging over measurements from multiple fields. However, these one-point statistics lack information on the correlations between pixels. For this reason, \citet{BarkanaLoeb_2008} have extended this formalism by analysing the one-point PDF of the difference in the differential 21cm brightness temperature measured at two points.

In this work, we present a new higher-order one-point statistic, to which we refer to as the local variance $\sigloc$. This local variance can be computed by dividing the total volume of a three-dimensional field $x(\bm{r})$ into N sub-volumes $\{V_\alpha\}_{1 \leq \alpha \leq N}$. Let us consider a sub-volume $V_\alpha$ centred on $r_\alpha$, described by the window function
\begin{equation}
\begin{aligned}
   & W_\alpha(\bm{r})=\prod_{j=1}^3 \Theta_j(\bm{r-r}_\alpha), & \Theta_j(\bm{r})= \left\{ \begin{aligned}
    & 1 & \mathrm{if} \ r_j\leq L_j/2,  \\
    & 0 & \mathrm{if} \ r_j > L_j/2,\\
    \end{aligned} \right.
\end{aligned}
\end{equation}
where $L_j$ is the length of each of the three sides of the sub-volume. Then the local variance is defined by:
\begin{equation}
    \sigloc^2  \equiv \sum_{\alpha=1}^N \left[ \int \mathrm{d}^3\bm{r}\ W_\alpha(\bm{r-r}_\alpha)\, x(\bm{r})  \right]^2 - \bar{x}^2,
\end{equation}
with $\bar{x}$ being the expectation value of the field. In other words, it is the variance of the distribution of the means of the sub-volumes. In this work, we mostly focus on the case where the sub-volumes considered are slices cut through the cube. It is clear from this definition that this estimator is based on sample variance and will be zero for a sufficiently large data cube. However, for smaller fields, it will provide us with information about the morphology imprinted in the field $x(r)$, as it -- contrary to usual one-point statistics, encompasses information on the correlations between pixels. It is interesting to note that previous works have also considered estimators computed in sub-volumes of data: for example, \citet{ChiangWagner_2015,GiriD'Aloisio_2019} investigate the power spectrum of sub-volumes, also referred to as the position-dependent power spectrum. However, in contrast to these approaches, our estimator will benefit from its simplicity.

We fully introduce the local variance and give its phenomenological definition in Sec.~\ref{sec:methods}. In Sec.~\ref{sec:results}, we apply this statistic to a range of simulations, that we describe in Sec.~\ref{sec:sims}, and find that its evolution with redshift is a good tracer of the ionisation history and topology, even when including observational effects, such as thermal noise and telescope resolution. We conclude in Sec.~\ref{sec:conclusions}.
In the following, all distances are in comoving units and the cosmology used is the best-fit cosmology derived from Planck 2015 CMB data \citep{planck_2015_overview}: $h = 0.6774$, $\Omega_\mathrm{m} = 0.309$, $\Omega_\mathrm{b} h^{2} = 0.02230$, $Y_\mathrm{p} = 0.2453$, $\sigma_8 =0.8164$ and $T_\mathrm{CMB} = 2.7255~\mathrm{K}$. We use interchangeably the terms filling fraction and ionisation level to describe the mean of ionisation fields $\bar{x}_e$.

\section{Simulations}
\label{sec:sims}

\begin{figure*}
    \centering
    \includegraphics[width=0.6\textwidth]{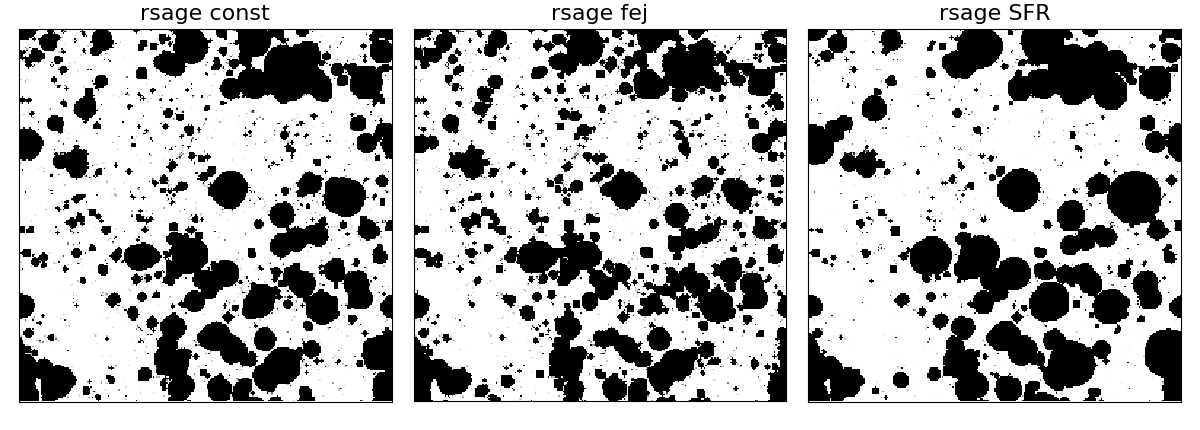}\\
    \includegraphics[width=0.6\textwidth]{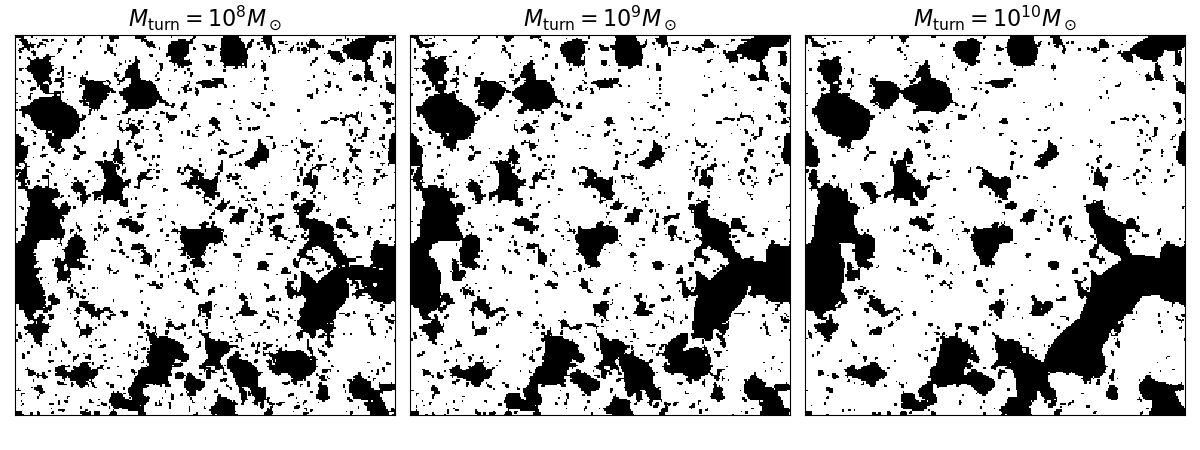}
    \caption{Binary ionisation fields cut through the three \texttt{rsage} simulations (upper panels) and the three 21CMFAST runs (lower panels) described in Sec.~\ref{sec:sims} at $\bar{x}_e = 0.30$, illustrating how different physics lead to a different reionisation morphology.}
    \label{fig:snapshots}
\end{figure*}

We applied our analysis to two types of simulations in order to check that our results are robust against different ways of modelling reionisation.
First, we consider the \texttt{rsage} simulations \citep{seiler_2019_fesc}, which are based on a $N$-body simulation with $2400^3$ dark matter (DM) particles and a side length of $160~\mathrm{Mpc}$. A modified version of the Semi-Analytic Galaxy Evolution (SAGE) model \citep{croton_2016}, which accounts for delayed supernovae feedback and radiative feedback, describes the evolution of the galaxies and their properties within the simulation. The ultraviolet background (UVB) is generated with the semi-numerical code \texttt{cifog} \citep{cifog,scifog_anne}. \texttt{cifog} also follows the time and spatial evolution of the ionised hydrogen regions in the simulation box.
Three different prescriptions for the escape fraction of ionising photons from galaxies into the IGM, $\fesc$, cover the physical plausible parameter space: In \texttt{rsage const}, $\fesc$ is considered to be constant regardless the redshift and  properties of the galaxies. Its value is fixed to $20\%$ \citep{robertson_2015}. In \texttt{rsage fej}, $\fesc$ scales with the fraction of gas ejected from each galaxy, $f_\mathrm{ej}$. In \texttt{rsage SFR}, $\fesc$ scales with the star formation rate of each galaxy, resulting in $\fesc$ effectively scaling with halo mass. These different ionising properties result in a different morphology of the ionisation fields, with \texttt{rsage SFR} exhibiting the largest ionised bubbles at a given global ionisation fraction $\bar{x}_e$. This is illustrated in the upper panels of Fig.~\ref{fig:snapshots}, which show the binary ionisation fields of the three simulations at $\bar{x}_e=0.3$, when the simulations are $30\%$ ionised. Since the three \texttt{rsage} simulations have the same underlying DM distribution and have been tuned to reproduce the Planck optical depth, we find their ionisation histories to be very similar (see the upper right panel of Fig.~\ref{fig:res_xhII_slices}). However, due to the different descriptions of $\fesc$, they diverge slightly towards the end of the reionisation process, with \texttt{rsage SFR} reaching a fully ionised IGM by $\Delta z\simeq0.1$ earlier than \texttt{rsage fej}.

Secondly, we use the publicly available 21CMFAST simulation \citep{21cmFAST_2007,21cmFAST_2011}\footnote{Available at \url{https://github.com/21cmfast/21cmFAST}.}. 21CMFAST is a semi-numerical code using excursion-set formalism \citep{furlanetto_2004_excursion_set}: starting from a matter overdensity field, it assumes each cell to be ionised when the number of photons exceeds the number of baryons in the respective cell. 21CMFAST has multiple simulation parameters that can be varied to change the underlying physical model, which again can result in different reionisation histories and morphology. Here, we choose to vary the parameter $\Mturn$, the turnover mass, which corresponds to the minimum halo mass below which star formation is suppressed exponentially. During the EoR, $\Mturn=10^8M_\odot$ would correspond roughly to a virial temperature of $10^4~\mathrm{K}$. We generate three simulations, with the same dimensions and resolution as \texttt{rsage} simulations, and assume $\Mturn=10^8M_\odot$, $10^9M_\odot$ and $10^{10}M_\odot$, to which we refer as M8, M9 and M10 in the following, respectively. Their global ionisation histories can be seen in the lower-right panel of Fig.~\ref{fig:res_xhII_slices}. The higher number of sources in M8 lead to an earlier reionisation of the IGM than in the other two simulations. In M10, however, reionisation is delayed until haloes of sufficient mass have formed. Since more massive sources are also more efficient at ionising their surroundings, M10 has on average larger ionised regions than M8 and M9. This can be seen in the lower panels of Fig.~\ref{fig:snapshots}, which show the snapshots of the ionisation fields of M8, M9 and M10 at $\bar{x}_e=0.3$.

Extracting the ionisation fields from the differential 21cm brightness temperature maps remains difficult \citep[e.g.][]{MalloyLidz_2013,BeardsleyMorales_2015,DattaGhara_2016,GiriMellema_2018,MangenaHassan_2020} due to their contamination by instrumental effects and foregrounds \citep{Gluscevic_Barkana_2010,ChapmanAbdalla_2013,LiuShaw_2020}: hence, we need statistical tools that we can apply directly to these 21cm maps. For this reason we apply our new one-point statistics also to the differential 21cm brightness temperature $\delta T_\mathrm{b}$ cubes in the following. From the \texttt{rsage} neutral fraction, $x_\ion{H}{I}=1-x_e$, and baryon density, $\delta_\mathrm{b}$, cubes, we construct the corresponding $\delta T_\mathrm{b}$ fields following \citep{pritchard_loeb_2012}:
\begin{equation}
\label{eq:def_brightness_T}
\delta T_\mathrm{b} = 27 x_\ion{H}{I} \left( 1 + \delta_\mathrm{b} \right)  \frac{\Omega_\mathrm{b}h^2}{0.023} \sqrt{ \frac{0.15}{\Omega_\mathrm{m}h^2}} \sqrt{ \frac{1+z}{10} }   ~\mathrm{mK}.
\end{equation}
Here, we assume that X-rays have heated the gas sufficiently such that the spin temperature of the neutral hydrogen gas exceeds the CMB temperature considerably. Because this work is a proof-of-concept for the local variance, we choose to limit our analysis to results using this assumption. However, recent works have shown that the CMB temperature might actually not be negligible during reionisation \citep{HenekaMesinger_2020}. For the 21CMFAST simulations, the brightness temperature cubes are computed directly by the simulation package, which allows a more complete derivation than the approximation given in Eq.~\eqref{eq:def_brightness_T}, namely including velocity corrections. Since interferometric observations will only measure the fluctuations in the 21cm signal, we subtract each cube by its mean so that the 21cm cubes have a mean zero. In practice, an interferometer measures a mean-zero map for each frequency bin, which is equivalent to saying that each slice of this cube has a mean zero and would make the local variance vanish. Hence, we assume that using $N$ slices of a coeval cube at fixed redshift $z$ is equivalent to considering $N$ small patches of a larger field-of-view at fixed frequency. We leave a detailed investigation of this hypothesis for future work.

\section{Methods}
\label{sec:methods}

In order to build intuition for this new estimator, we first look at the ionisation fields of our simulations. We obtain the 3D variance of the ionisation field of a coeval cube extracted from the simulations, at a given redshift $z$ and global ionisation level $\bar{x}_e$, by computing:
\begin{equation}
\label{eq:def_sigwhole_numerical}
    \sigwhole(z)^2 = \frac{1}{N^3} \sum_{i,j,k=1}^N  x_{i,j,k}(z)^2 - \bar{x}_e^2 
\end{equation}
where the ionisation level of a cell $(i,j,k)$ can be either $x_{i,j,k}(z) = 0$ or $1$. It is possible to relate the variance $\sigwhole^2$ of a 3D field $x_e(\bm{r})$ to its 2-point correlation function (2-PCF) $\xi_2$ and, in turn, to its power spectrum $P(\bm{k})$\footnote{The definition of the 2-PCF yields
\begin{equation}
\label{eq:def_xi2}
    \xi_{2}(\bm{r})=\frac{1}{V} \int \mathrm{d}^3 \bm{s}\ x_e(\bm{s}) \,x_e(\bm{s}+\bm{r})=\frac{1}{(2\pi)^3} \int \mathrm{d}^3\bm{k} \, P(\bm{k})\, \mathrm{e}^{i\bm{k}\cdot \bm{r}},
\end{equation}
such that, for an isotropic and homogeneous field,
\begin{equation}
\label{eq:sigwhole_PS}
    \sigma^2_\mathrm{whole} + \bar{x}_e^2 = \frac{1}{V} \int \mathrm{d}^3 \bm{s} \ x_e(\bm{s})^2 =\xi_2(0) = \frac{1}{2\pi^2} \int k^2 \mathrm{d}k \, P(k).
\end{equation}}.
With this relation, we can estimate the variance of the EoR 21cm signal by measuring its power spectrum. In this context, \citet{Patil_2014} has used forecast LOFAR observations and the inferred redshift-evolution of $\sigwhole$ to constrain the midpoint and duration of reionisation. 
From a topological point of view, examining the evolution of $\sigma_\mathrm{whole}$ with $\bar{x}_e$ allowed \citet{watkinson_2014_1PCF} to differentiate between outside-in and inside-out scenarios of reionisation. However, in all the simulations analysed in this work, reionisation proceeds inside-out. As such, we find for our three \texttt{rsage} simulations only a $\sim 1\%$ deviation from the theoretical parabola which can be derived from the PDF of ionised pixels $P(x_e)$ given in Eq.~\eqref{eq:def_PDF_3D_pixels}:
\begin{equation}
\label{eq:sigma_whole}
\sigwhole^2 = \int (x_e-\bar{x}_e)^2\, P(x_e) \ \drm x_e = \bar{x}_e \left(1-\bar{x}_e \right).
\end{equation}
The results also hold for the 21CMFAST simulations and higher order cumulants, such as the skewness and the kurtosis. Therefore, the distribution of pixels throughout the whole box cannot differentiate between the simulations, as it mainly traces the reionisation history. In particular, it does not account for the correlations between pixels, and hence cannot track morphological differences.

\begin{figure*}
    \centering
    \includegraphics[height=5.cm]{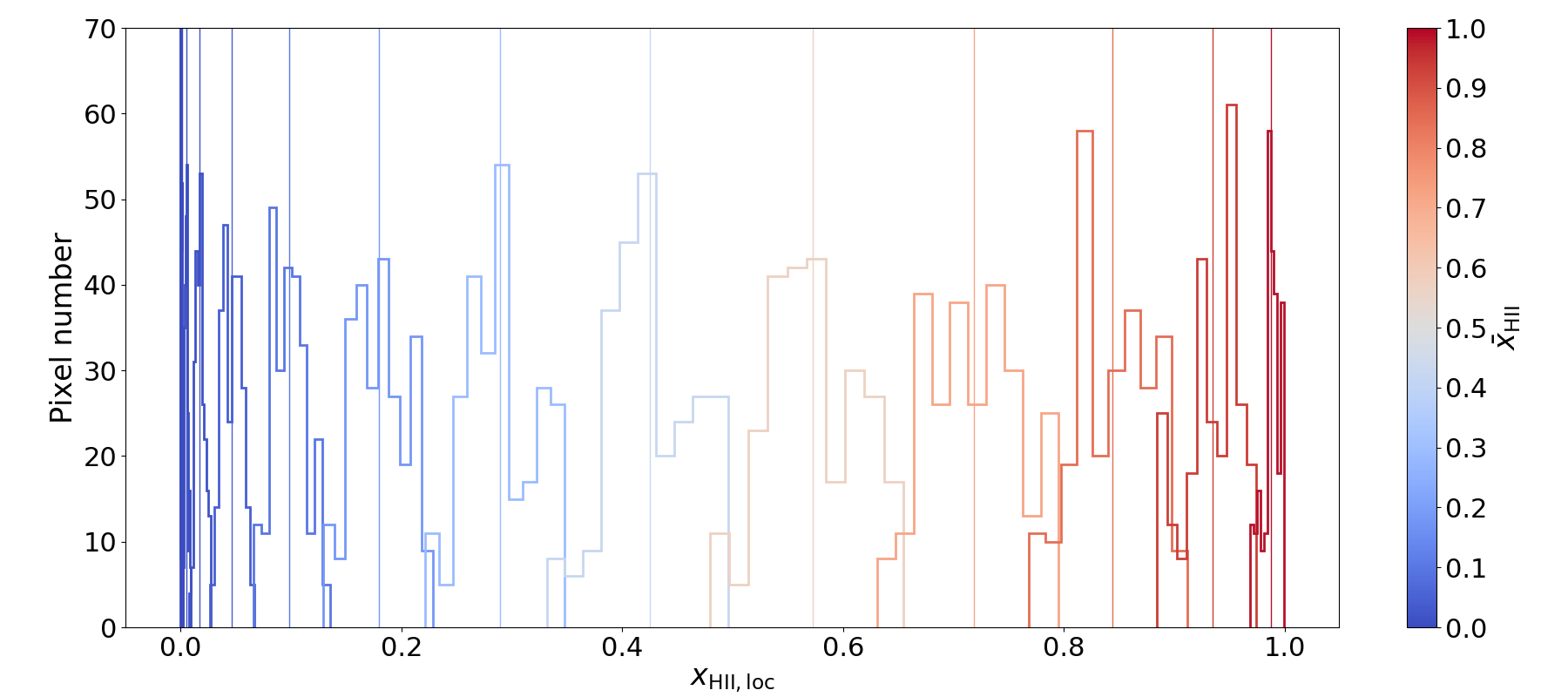} \includegraphics[height=5.cm]{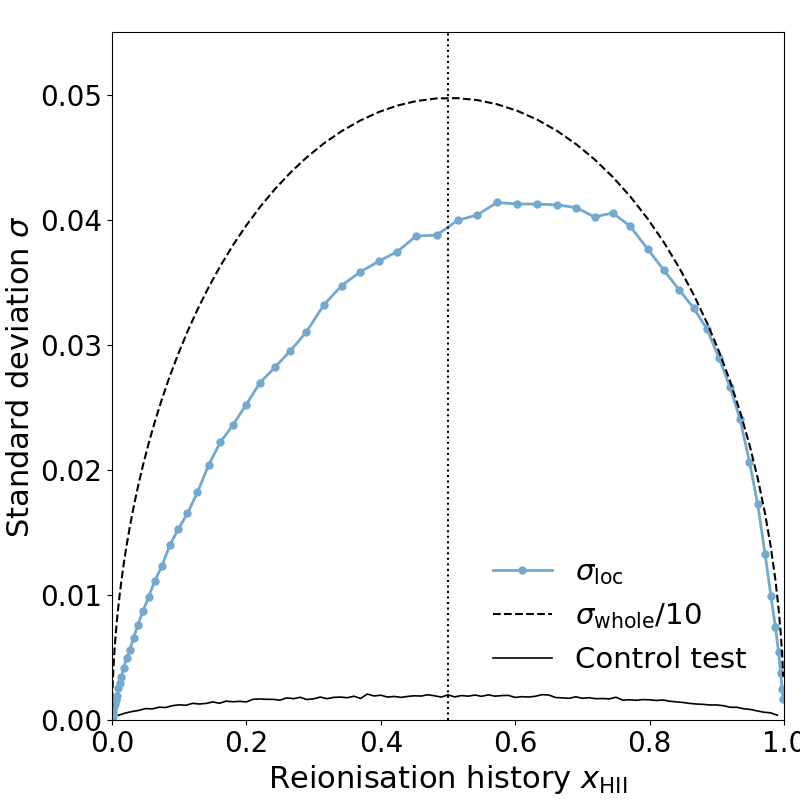}\\
    \caption{\textit{Left panel:} Distribution of the ionisation levels of the N slices that can be carved out of the \texttt{rsage const} simulation along one direction. Each colour corresponds to one of 12 snapshots taken on the range $6.02 \leq z \leq 14.63$, corresponding to different ionisation levels, represented as the solid vertical lines. \textit{Right panel:} Evolution of the standard deviations of each distribution with global reionisation history (blue solid line), compared to the standard deviation of the distribution of ionised pixels throughout the whole box (dashed line, divided by 10).}
    \label{fig:rsage_const_slices}
\end{figure*}

For this reason, we introduce the local variance, which is sensitive to the ionisation topology as it includes the small and large-scale correlations between points:
\begin{equation}
\begin{aligned}
    \sigloc(z)^2
    & = \frac{1}{N} \sum_{k=1}^N \left( \frac{1}{N^2} \sum_{i,j=1}^N  x_{i,j,k}(z) \right)^2 - \left( \frac{1}{N^3} \sum_{i,j,k}^N  x_{i,j,k}(z) \right)^2 \\
    & = \frac{1}{N} \sum_{k=1}^N \mu(k,z)^2 - \bar{x}_e(z)^2.
\label{eq:def_sigloc}
\end{aligned}
\end{equation}
Here $\mu(k,z)$ is the mean value of the $k^\mathrm{th}$ slice along the redshift direction\footnote{Because we use coeval cubes, we assume that the redshift does not change from one slice to the next. This is a reasonable assumption because of the relatively small size of the box ($L=160~\mathrm{Mpc}$).}.
We explain the derivation of this statistic by considering the three-dimensional binary ionisation field of the \texttt{rsage const} simulation. The ionisation field is a cube with $N=256$ cells on a side, each cell having a width of $\Delta x = 0.625~\mathrm{Mpc}$. Snapshots every $10~\mathrm{Myrs}$, tracking the reionisation process, are available. For each snapshot, we compute the average value $\mu$ of each of the $N$ slices, each having a width of one cell along a chosen direction. Here, we assume this direction to be the redshift -- or frequency -- direction. The resulting distribution of $N$ filling fractions $\{ \mu \}_{0 \leq k < N}$ is centred around the filling fraction of the whole 3D box at the respective redshift, $\bar{x}_e(z)$. The left panel of Fig.~\ref{fig:rsage_const_slices} shows these distributions for redshifts $6 \leq z \leq 15$. At the beginning of reionisation, the distribution is very narrow but widens as reionisation progresses and ionised bubbles grow, tracing the underlying clustered galaxy population and the increasing correlation between pixels. At the end of reionisation, the distribution is again very narrow as most pixels are fully ionised. In the right panel of the figure, we summarise our results by plotting the evolution of the standard deviation of these distributions $\sigma_\mathrm{loc}$ as a function of the global ionisation level (blue solid line). 

As outlined in our motivation, we can see from Eqs.~\eqref{eq:def_sigwhole_numerical} and \eqref{eq:def_sigloc}, that the local variance $\sigloc$ describes the morphology of the considered field more accurately than the ordinary 3D variance $\sigwhole$, as it includes the variance of each slice. Indeed, if we consider $\mathrm{Var}(k)$ the variance of the $k$-th slice, then $\mathrm{Var}(k) - \mu(k) = \sum^N_{i,j}x_{i,j,k}^2/N^2$, and it is easy to see that
\begin{equation}
    \sigloc^2 = \sigwhole^2 - \frac{1}{N} \sum_{k=1}^N \mathrm{Var}(k).
\end{equation}
%For a Gaussian random field, where such correlations do not exist, the local variance can be related to $\sigwhole$,
% \begin{equation}
%     \sigloc_\mathrm{,GRF} = \sigwhole_\mathrm{,GRF}/N,
% \end{equation}
% where the slices have dimensions $N^2$. This relation will be particularly useful when we add thermal noise to the 21cm fields in Sec.~\ref{subsec:4_noise}. 
Furthermore, we note that since the 3D variance can be expressed in terms of the 2-PCF given in Eq.~\eqref{eq:sigwhole_PS}, the local variance is also given by
\begin{equation}
\label{eq:sigloc_vs_PS}
    \sigma_\mathrm{loc}^2 =\frac{1}{L} \int \mathrm{d}r\, \mu^2(r) =  \frac{1}{2\pi} \int \mathrm{d}k \, P_\mu(k),
\end{equation}
where $\mu(r)$ is the mean of the slice located at $r$ and $P_\mu(k)$ is the power spectrum of the 1D distribution of means $\{ \mu(r) \}_{r \leq L}$. $P_\mu(k)$ corresponds to the 3D power spectrum of the field, when only the modes along the frequency direction in Fourier space are kept and are rescaled by the area of the observational window in the sky plane: $P_\mu(k)=P(\bm{k})/L^2$ for $\bm{k}=(0,0,k_z)$. 
Selecting such modes can be done by using a specific window function, for example a Bessel function \citep{munoz_2020_cosmic_variance}.

\section{Results}
\label{sec:results}

\subsection{Understanding the local variance}
\label{subsec:sigloc_rsage_const}

In this section, we analyse the evolution of the local variance of the \texttt{rsage const} simulation. In order to understand the impact of the ionisation fraction and density distributions on our estimator, we first discuss the local variance of the ionisation fraction fields before we analyse the local variance of the differential 21cm brightness temperature maps. For clarity, we add a superscript to $\sigloc$, describing the field considered: $\siglocion$ for the ionisation field, $\siglocbright$ for the brightness temperature field.

We show the local variance of the ionisation field $\siglocion$ of the \texttt{rsage const} simulation as a function of its mean in the right panel of Fig.~\ref{fig:rsage_const_slices}. For comparison, we also plot the results for the scaled 3D variance, $\sigwhole/10$. The dotted vertical line indicates the midpoint of reionisation at $\bar{x}_e=0.50$, where $\sigma_\mathrm{whole}$ is maximum (see Eq.~\eqref{eq:sigma_whole}).
On the other hand, $\siglocion(\bar{x}_e)$ (blue line) is slightly distorted compared to $\sigma_\mathrm{whole}$ and reaches its maximum around $\bar{x}_e\simeq0.60$. The location of the maximum indicates the moment when ionised and neutral regions are the largest, which will depend on the large-scale structure of the ionisation fields. We discuss this in more detail in the next Section when we compare different ionisation morphologies. To confirm the physical origin of this signal, we compute the local variance of a control test, consisting of a 3D box of the same resolution and size as our simulations, but randomly filled with ionised pixels to reach the considered ionisation level. Such a field will contain none of the correlations we aim at probing with the local variance and will actually be analogous to a box with white noise power spectrum. It can also be seen as a field made of many uncorrelated very small bubbles, which is close to the ionisation field at the very beginning of reionisation (see next paragraph). The resulting variance, close to zero and largely insignificant compared to what was obtained for the simulations, is shown as the black solid line in the right panel of Fig.~\ref{fig:rsage_const_slices}. 

\begin{figure}
    \centering
    \includegraphics[height=7.5cm]{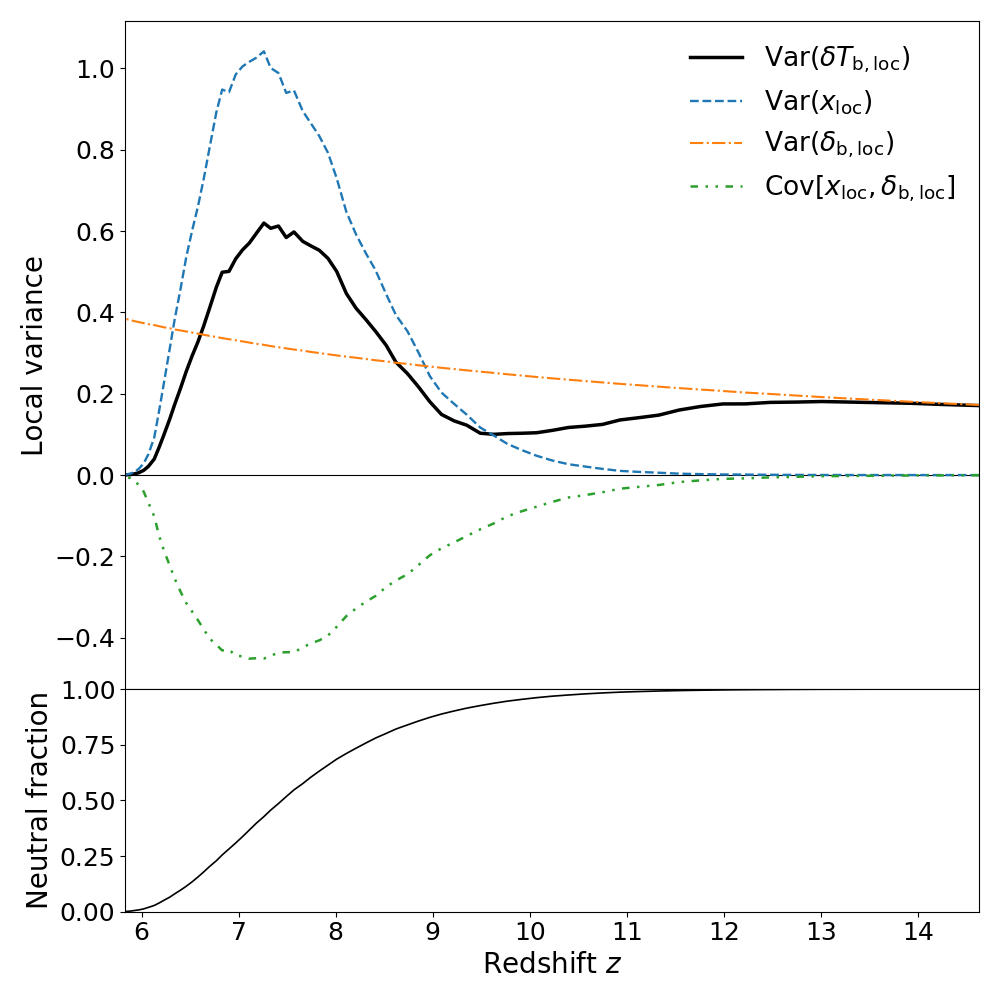}
    \caption{Contributions to the local variance of the $\delta T_b$ field in the \texttt{rsage const} simulation (upper panel) and its reionisation history (lower panel). }
    \label{fig:varloc_tb_contributions}
\end{figure}

In Fig.~\ref{fig:varloc_tb_contributions}, we show the local variance of the $\delta T_\mathrm{b}$ field of the \texttt{rsage const} simulation as a function of redshift (thick solid line) along with the local variances of the $\ion{H}{I}$ and $\delta_\mathrm{b}$ fields and the covariance of the distributions of means for the ionisation field (${\xloc}$) and the 21cm brightness temperate field (${\delta_\mathrm{b,loc}}$). Because of correlations, the exact expression of $\siglocbright$, given in App.~\ref{app:dTb_var_calc}, does not equal the sum of the aforementioned three elements, but they are useful to understand the behaviour of $\siglocbright$. 
Overall, the redshift-evolution of the local variance $\siglocbright(z)$ traces the reionisation history and is similar to the one observed in \citet{Patil_2014} for the 3D variance. At high redshift, before the onset of reionisation ($z\gtrsim12$), the variance across slices comes from the underlying density field.
As the first sources start ionising neutral hydrogen, zero pixels appear on the Gaussian $\delta T_\mathrm{b}$ background ($z\simeq9-12$), and homogenise the distribution \citep{Gluscevic_Barkana_2010,BittnerLoeb_2011}, leading to a dip in the variance at $z\simeq9.5$. At this point, the distribution of ionised regions is similar to the one of the control test, considered above, hence the small amplitude of the local variance.
As reionisation proceeds, zero (ionised) pixels start tracing the reionisation morphology. As their number increases compared to the number of warm (neutral) pixels (which still follow a Gaussian distribution), $\siglocbright$ starts mostly tracing the ionisation field and increases until reaching its maximum around the midpoint of reionisation, when both ionised and neutral regions are the largest. As more and more pixels are ionised, the global brightness temperature approaches zero and so does the local variance.
A similar redshift evolution has been observed in the skewness of pixel distributions within two-dimensional brightness temperature maps \citep{HarkerZaroubi_2009} and can be recovered using a combination of analytical functions \citep{IchikawaBarkana_2010,Patil_2014}. However, these theoretical functions do not say much about the time and spatial distribution of ionised regions.

\subsection{Comparing simulations}
\label{subsec:comparison_sigloc}

\begin{figure*}
    \centering
    \includegraphics[height=5cm]{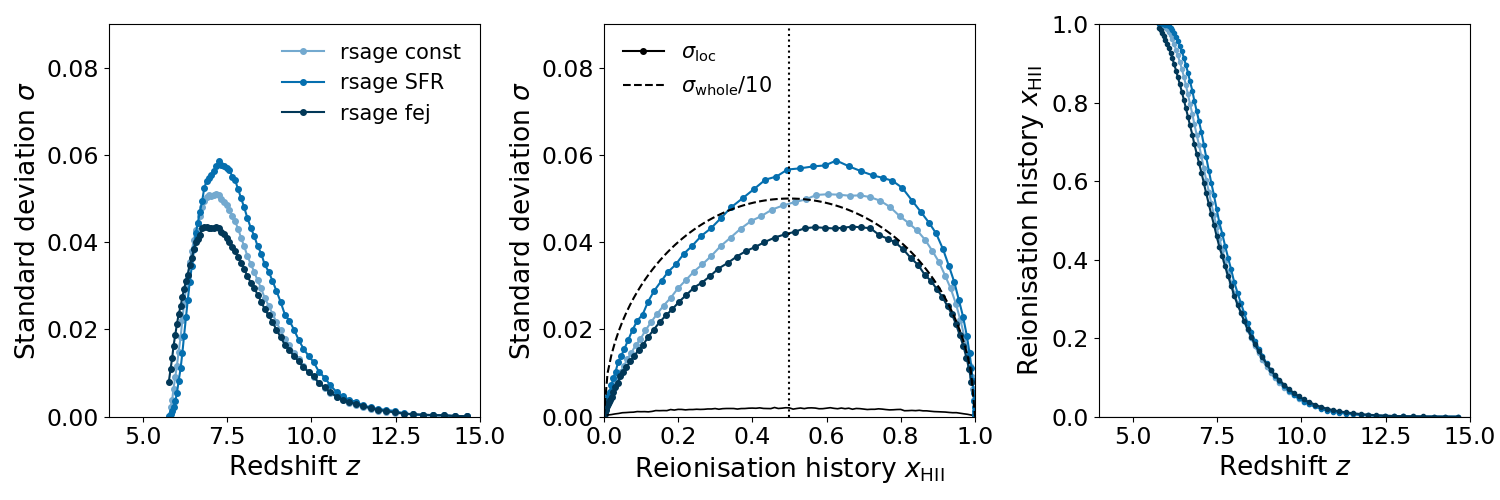}
    \includegraphics[height=5cm]{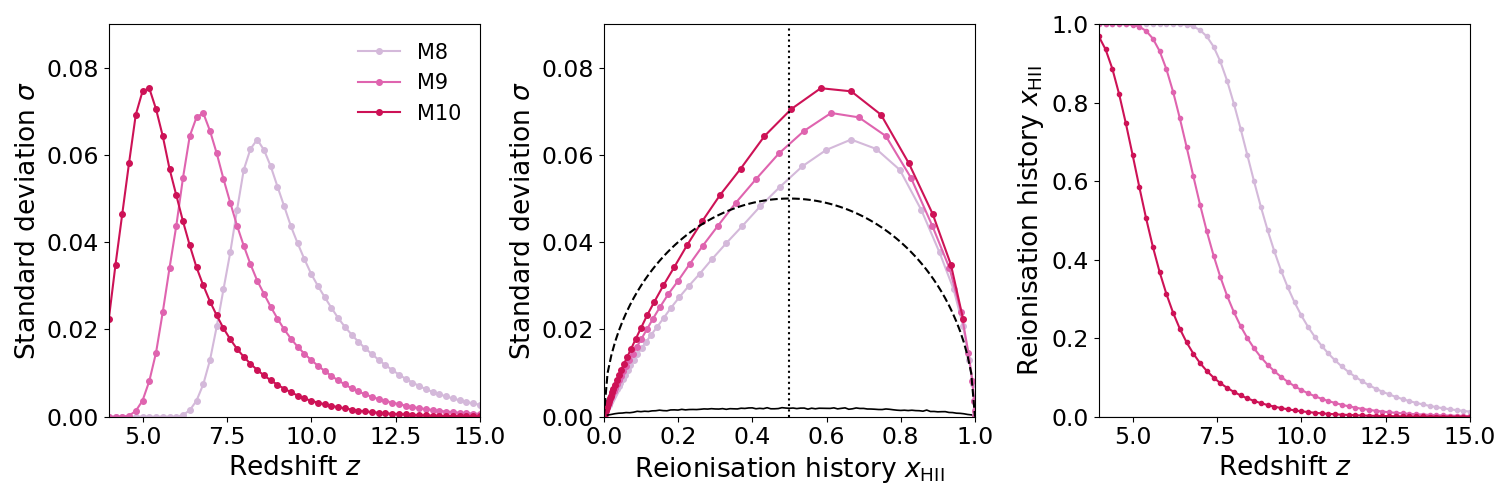}
    \caption{Evolution of the standard deviation on the distribution of means measured in the set of N slices that can be carved out of the ionisation fields of simulations along one direction with redshift (left panel) and ionisation level (middle panel). \textit{Right panel:} Corresponding reionisation histories. Results for the three \texttt{rsage} simulations (upper panels) and the three 21CMFAST runs (lower panels) are compared.}
    \label{fig:res_xhII_slices}
\end{figure*}

In order to understand how the ionisation morphology affects the characteristic features in $\sigloc(z)$, which are the amplitude and ionisation fraction at which $\sigloc(z)$ reaches its maximum, we compare $\siglocion(z)$ for all the simulations described in Sec.~\ref{sec:sims}. Results for \texttt{rsage} and 21CMFAST are shown in the upper and lower panels of Fig.~\ref{fig:res_xhII_slices}, respectively.

As we can see from the top right panel, the three \texttt{rsage} simulations show similar reionisation histories, and, therefore, the approximate locations of the minima and maxima of the local variance are similar. However, they differ in their amplitudes. Here, in contrast to what we have found for $\sigwhole$, there is a clear difference between the three \texttt{rsage} simulations: for example, at $\bar{x}_e=0.50$, the local variances of \texttt{rsage fej} and \texttt{rsage SFR} are about $20\%$ below and $20\%$ above the one of \texttt{rsage const}, respectively. We find more variance between the \texttt{rsage SFR} slices, since the simulation exhibits larger ionised regions \citep{seiler_2019_fesc}. This can be understood as follows. Consider an ionisation field at a given global ionisation level $\bar{x}_e$. If the field is made of a few large ionised bubbles and we cut a slice through the box, we are more likely to pick up a large ionised region that will bias the filling fraction of the slice $\mu$ towards values larger than $\bar{x}_e$. Conversely, if the field is made of many small ionised regions, such as in the \texttt{rsage fej} simulation, the slices cut through the box are more likely to have similar $\mu$ values and $\siglocion$ will be lower. 

We confirm these findings and extend our understanding of how the characteristics of $\siglocion$ depend on the reionisation morphology by analysing the results we obtain for the three 21CMFAST simulations that differ in their reionisation history and morphology (see Fig.~\ref{fig:snapshots}).
When computing the local variance of the 21CMFAST ionisation fields, we find that, similarly to \texttt{rsage}, at a given ionisation level, the simulation with the on average largest ionised regions, M10, yields the largest $\siglocion$ values. This is in agreement with the findings of \citet{Gluscevic_Barkana_2010}, who already noticed that if the ionising sources lie in more massive haloes in one simulation than another, the impact on the 3D pixel distribution is noticeable as the ionised regions are larger and more scarce at the same global ionisation fraction. Since the three 21CMFAST simulations exhibit not only different ionisation morphology but also different ionisation histories, the redshift-evolution of $\siglocion$ varies from one simulation to the other in addition to its variations in amplitude. This result implies that measuring the local variance of a field will help us to constrain the reionisation history.

\begin{figure*}
    \centering
    \includegraphics[width=.65\textwidth]{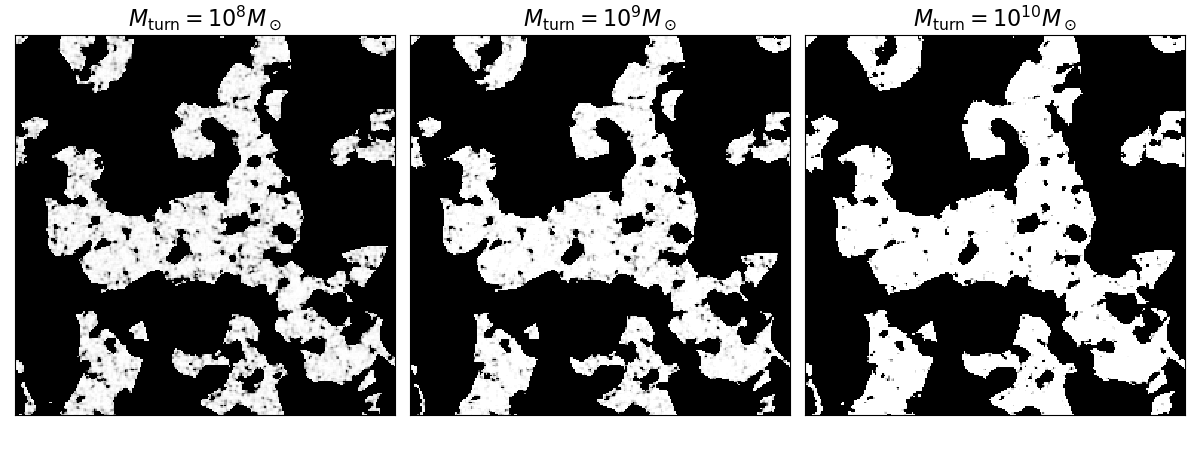}
    \caption{Snapshots of the ionisation field at the maximum of the local variance for the M8, M9 and M10 simulations, corresponding to different redshifts and different ionisation levels: $z=8.4$ and $\bar{x}_e=0.67$ for M8, $z=6.7$ and $\bar{x}_e=0.65$ for M9, and $z=5.2$ and $\bar{x}_e=0.58$ for M10.}
    \label{fig:snapshots_sigmax}
\end{figure*}

Similar to the \texttt{rsage} simulations, the maximal local variance is also reached around the reionisation midpoint for the three 21CMFAST runs. In general, this is expected to happen when the derivative of the global signal with respect to redshift is maximal \citep{munoz_2020_cosmic_variance}. During reionisation, we expect the signal to be maximal when both ionised and neutral regions are the largest. Applying the bubble size algorithm granulometry \citep{KakiichiMajumdar_2017} to the \texttt{rsage} simulations, we find this to be the case for all three simulations, at $\bar{x}_e\sim0.6$ \citep{hutter_2020_bispectrum}, which also corresponds to the measured maximum of the local variance in the simulations. Indeed, we find the maximum to be located at $z = 8.4 \pm 0.1$, $6.7\pm0.1$ and $5.2\pm0.1$, corresponding to an ionisation level of $\bar{x}_e=0.67\pm 0.02$, $0.65\pm 0.02$ and $0.58\pm 0.02$ for the M8, M9 and M10 simulations, respectively. Interestingly, this result holds when computing the local variance of a toy model, made of randomly located fully ionised bubbles. All the bubbles have the same initial radius and are grown by increasing the radius one pixel at a time until the whole box is ionised (details on this toy model can be found in App.~\ref{app:toy_model}). In these toy models, the maximum local variance is always reached when the box is about $60\%$ ionised, although slightly sooner when the initial bubble radius is larger.  %The exact timing will depend not only on the ionising properties of the sources -- for example, on the value of $\Mturn$ or $\fesc$ --, but also on the clustering of the underlying density field. Indeed, since the three \texttt{rsage} simulations reach $60\%$ ionisation at roughly the same redshift, $z \sim 7$, the small (less than $5\%$) difference in their local variance maxima seems to purely originate from the difference in the ionising properties of the sources. In contrast, Table~\ref{tab:sigmax} shows that the 21CMFAST M8 simulation reaches its maximum local variance at a higher redshift, $z=8.4\pm0.1$, than M10, $z=5.2\pm 0.1$. However, this difference is not only due to reionisation occurring earlier in M8 but also the ionisation morphology: at $z=8.4$, $67\%$ of M8 is ionised, whereas at $z=5.2$, only $58\%$ of M10 is. 
In Fig~\ref{fig:snapshots_sigmax}, we show snapshots of the ionisation fields of the three 21CMFAST simulations at the redshift when they reach the maximum local variance. Interestingly, despite these snapshots corresponding to different ionisation levels and redshifts, they have their largest ionised regions in common. This indicates that, in contrast to its amplitude, the location of the maximum of the local variance is purely sensitive to the large-scale structure of the ionisation field. As the total number of ionising photons decreases in M8, M9 and M10, respectively, the large-scale ionised regions will reach their maximum size at lower redshifts; but due to the different ionising emissivity distributions across sources, the redshifts where the local variance becomes maximal will correspond to a different ionisation level of the box. For example, in Fig~\ref{fig:snapshots_sigmax}, we see that the neutral regions are filled with many small ionised regions in M8, increasing the overall ionisation level of the simulation to a higher one than in M10 at the redshift of maximal local variance.

\begin{figure}
    \centering
    \includegraphics[height=6cm]{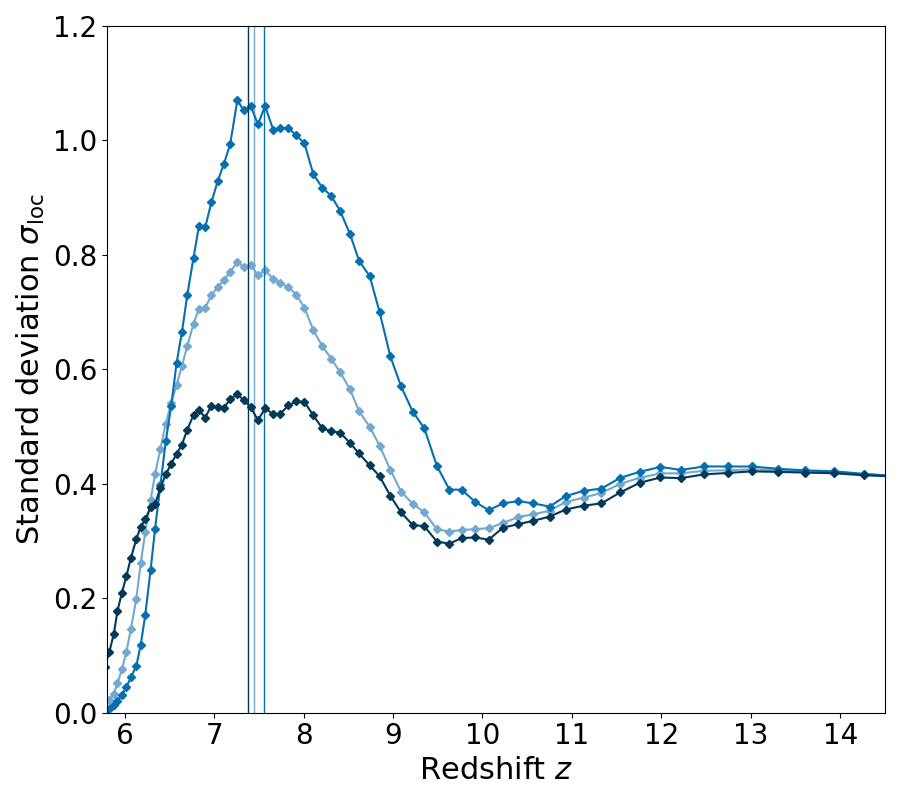}
    \includegraphics[height=6cm]{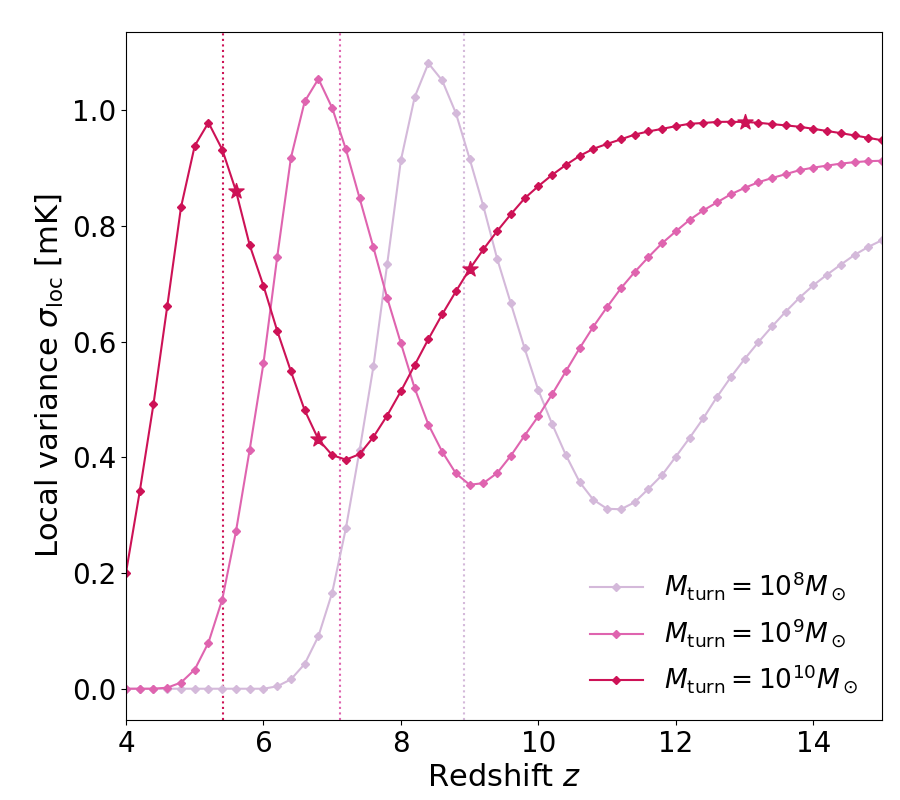}
    \caption{Local variance of the 21cm brightness temperature fields from the \texttt{rsage} (upper panel) and 21CMFAST (lower panel) simulation, as a function of redshift. Vertical dotted lines show the ionisation midpoint of the simulation of the corresponding colour.}
    \label{fig:sigloc_21cm}
\end{figure}

We have seen that differences in the ionisation morphology across simulations translate into differences in the amplitude of $\siglocion(z)$, and differences in reionisation histories into translations in redshift. We now turn to the differential 21cm brightness temperature fields which, as we have seen in the previous Section, encompass additional information from the ionisation morphology. Figure~\ref{fig:sigloc_21cm} shows the redshift-evolution of $\siglocbright$ with redshift for the three \texttt{rsage} simulations (upper panel) and the three 21CMFAST runs (lower panel). We briefly note that the three \texttt{rsage} simulations show the same local variance at high redshifts ($z\gtrsim12$), because $\siglocbright$ is governed by the same underlying density field. However, as $\siglocbright$ becomes sensitive to the ionisation field, its shape traces the reionisation history, while its amplitude is sensitive to the reionisation morphology: as observed for the ionisation field, here, the \texttt{rsage SFR} simulation, which has the largest ionised regions on average, exhibits the largest $\siglocbright$. However, this is not true for the 21CMFAST simulations: in contrast to what is expected, M10 exhibits the lowest amplitude in $\siglocbright$ during reionisation. This is because M10 reionises later than M8, when the density field is more heterogeneous and the local variance of the density field larger; such that the anti-correlation between $\delta_b$ and $x_\ion{H}{I}$ is stronger, adding negative signal to the local variance of the overall brightness temperature field (see Fig.~\ref{fig:varloc_tb_contributions}). The pre-factor of Eq.~\ref{eq:def_brightness_T}, which is proportional to $\sqrt{1+z}$, also contributes to enhancing the signal at higher redshifts and hence to M8 showing a higher amplitude at its maximum because it is reached at a higher redshift. This also explains why M10 exhibits the shallowest dip at $z\simeq7$ of the three simulations during cosmic dawn. Finally, the larger amplitudes of the local variance seen at high redshift ($z\gtrsim12$ for M8, $z\gtrsim8$ for M10) is found to be related to the extra terms included in the 21CMFAST derivation of the 21cm brightness temperature compared to the simplified expression given in Eq.~\ref{eq:def_brightness_T}, namely the velocity field.

\subsection{Observational effects}
\label{subsec:4_noise}

Many limitations related to the nature of instruments are expected to complicate the observation of the 21cm signal from reionisation. For this reason, we consider the effects of thermal noise and angular resolution smoothing on our $\delta T_\mathrm{b}$ maps and subsequent measurements of the local variance. In the following, we consider the performance characteristics of SKA1-\textit{Low} \citep{BraunBonaldi_2019}, in an optimistic and a pessimistic case, corresponding respectively to a maximum baseline of $b_\mathrm{max}=65~\mathrm{km}$ and $2~\mathrm{km}$. In both cases, the total effective collecting area of $A_\mathrm{tot}\sim 10^5~\mathrm{m}^2$ that is frequency dependent. Indeed, the Australian interferometer will consist of about $2\times 10^6$ dipoles, gathered in 512 stations with 24 tiles per station. Each individual dipole will have an effective area of $\lambda_{21}^2/3$, with $\lambda_{21}$ being the redshifted 21cm wavelength. 

In order to apply the appropriate SKA1-\textit{Low} angular smoothing to our $\delta T_\mathrm{b}$ maps, we convolve each simulation cube (corresponding to a given redshift $z$) by a Gaussian kernel with a FWHM of $\theta(z) \, d_c(z)$, with $d_c(z)$ being the comoving distance to redshift $z$ and 
\begin{equation}
   \theta(z) =  1.22\times \frac{\lambda_{21}(z)}{b_\mathrm{max}} 
\end{equation}
the angular resolution of the telescope.
Because of the size of the SKA1-\textit{Low} array, its angular resolution will be very high: It will range from $0.15~\mathrm{Mpc}$ at $z=4$ to $0.66~\mathrm{Mpc}$ at $z=15$, which is smaller than the resolution of our simulation grids ($\Delta x = L/N = 0.625~\mathrm{Mpc}$) at all redshifts of interest. Consequently, the smoothed and original maps are very similar, and so will be the resulting $\sigloc$. 

\begin{figure}
    \centering
    \includegraphics[height=6cm]{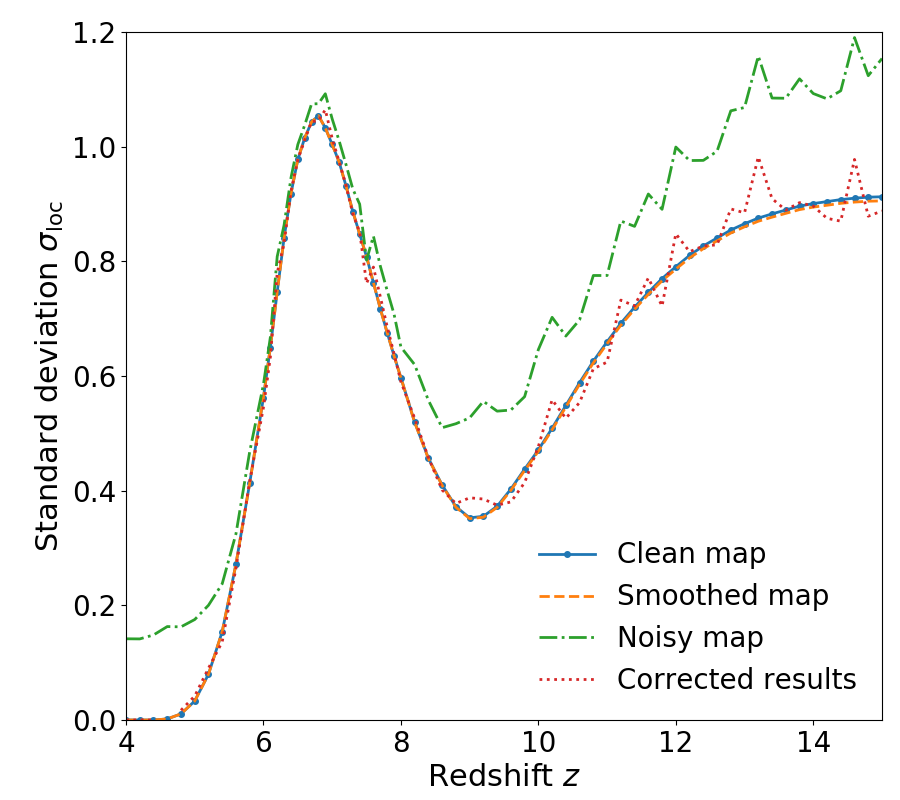}
    \caption{Local variance of the brightness temperature maps of the M9 21CMFAST simulation, for a clean map (solid blue line), a map smoothed to SKA1-\textit{Low} angular resolution for the optimistic case ($b_\mathrm{max}=65~\mathrm{km}$, dashed orange line), and a noisy map (dash-dotted green line). See text for details.}
    \label{fig:with_noise_mturn9}
\end{figure}

We then generate thermal random noise by drawing a noise value $n_i$ from a Gaussian distribution with a variance $\sigma_\mathrm{th}^2$ for each pixel, with the variance given by \citep{watkinson_2014_1PCF}:
\begin{equation}
\label{eq:noise_variance}
\begin{aligned}
\sigma_\mathrm{th}^2(z) =  2.9~\mathrm{mK}&  \times \left(\frac{10^5 ~\mathrm{m}^2}{A_\mathrm{tot}}\right)  \left( \frac{10~\mathrm{arcmin}}{\Delta \theta}\right)^2 \\
& \times\left(\frac{1+z}{10}\right)^{4.6} \sqrt{ \frac{1~\mathrm{MHz}}{\Delta \nu} \frac{100~\mathrm{h}}{t_\mathrm{int}}}.
\end{aligned}
\end{equation}
Here, $\Delta \nu$ is the frequency resolution of the experiment, which we match for computational efficiency to the resolution of the simulation $\Delta x$ according to $\Delta \nu =H_0 \nu_0 \sqrt{\Omega_\mathrm{m}} \Delta x /[c \sqrt{(1+z)}]$ with $H_0$ being the Hubble constant and $\nu_0$ the rest-frame 21cm frequency. SKA1-\textit{Low} is expected to have a much better frequency resolution than the comoving cell size of $0.625~\mathrm{Mpc}$ used in this work, with a channel width of $5.4~\mathrm{kHz}$ at a nominal frequency of $100~\mathrm{MHz}$ \citep{BraunBonaldi_2019}. A thinner resolution will be beneficial to local variance analyses (see App.~\ref{subsec:discussion_box_size}). We consider an observation time of $t_\mathrm{int}=1000~\mathrm{hrs}$. Using the variance given in Eq.~\ref{eq:noise_variance}, we add a noise value to each pixel of the smoothed 21cm brightness temperature map and compute $\sigloc$ for the resulting coeval cubes. 

\begin{figure*}
    \centering
    \includegraphics[height=6cm]{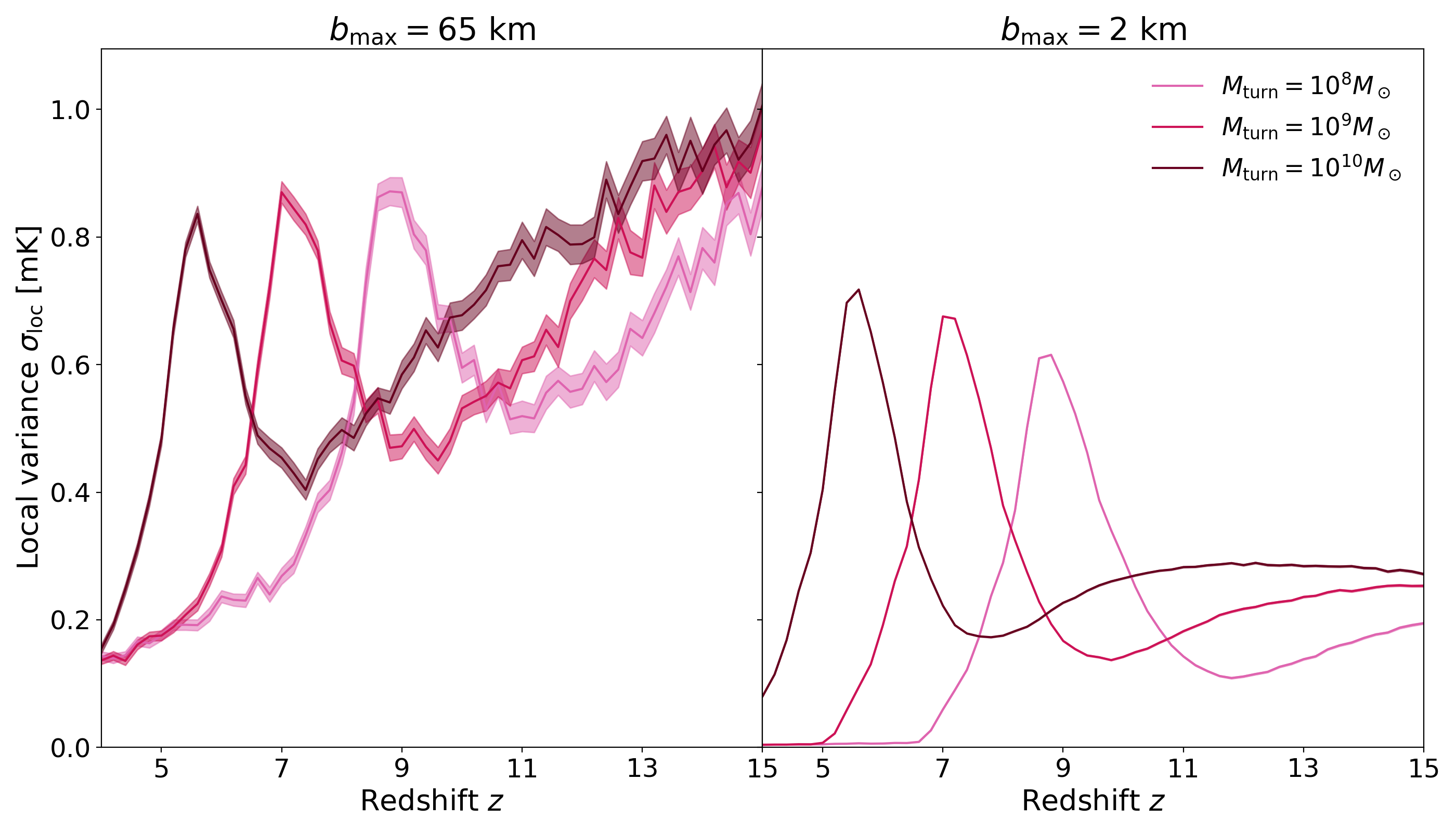}
    \caption{Local variance of the 21cm brightness temperature maps of the 21CMFAST simulations, after applying telescope resolution smoothing and adding telescope noise, for an optimistic case ($b_\mathrm{max}=65~\mathrm{km}$, left), and a pessimistic case ($b_\mathrm{max}=2~\mathrm{km}$, right). The shaded regions correspond to the standard deviation of the $\sigloc$ values obtained from 100 different realisations of the thermal noise.}
    \label{fig:comparison_bmax_21cmfast}
\end{figure*}

In Fig.~\ref{fig:with_noise_mturn9}, we show the local variance computed from the clean, smoothed and noisy map extracted from the M9 simulation, for the optimistic case. We see that, because $\sigloc$ is based on variance information and, therefore, not sensitive to the absolute value of the 21cm differential brightness temperature, the variance information is still accessible in both smoothed and noisy maps, despite the amplitude of the noise being comparable to the one of the cosmological signal. On the range of redshifts corresponding to the bulk of the reionisation process ($6.4 \leq z \leq 8.2 $ for $0.2 \leq x_e \leq 0.8$), the signal-to-noise ratio is above one, reaching its maximum of 3.1 when the signal is maximum.
The noise variance, computed using Eq.~\eqref{eq:noise_variance}, increases with increasing redshift, which provides an explanation for the rough edges of $\sigloc$ in the noisy maps at $z>8$. In fact, the noise and the cosmological signal being uncorrelated, we have%we have seen earlier that, for a Gaussian random field, $\sigloc=\sigwhole/N$ where the slices have dimensions $N^2$. Hence, 
\begin{equation}
    \label{eq:sigloc_smoothed_noisy}
    \sigma_\mathrm{loc,smoothed}^2 \simeq \sigma_\mathrm{loc,noisy~map}^2 - \sigma_\mathrm{loc,noise}^2.
\end{equation}
Subtracting the two local variances, we obtain the results shown as the dotted line in Fig.~\ref{fig:with_noise_mturn9} and refer to these as the corrected results in the following. In these corrected results, the noise bias has been removed, and the measured local variance is much closer to the local variance values obtained from the clean smoothed maps than from the uncorrected one.  Although Eq.~\eqref{eq:sigloc_smoothed_noisy} is an approximation and overlooks potential correlations between the cosmological signal and the noise within a slice, the good match between $\sigloc$ derived from clean maps and from corrected maps shows that their contribution is sufficiently small to justify this approximation.
In the optimistic case, the noise variance, and the associated fluctuations at high redshift, still remain but the location of the maximum of the local variance can be recovered. 

In Fig.~\ref{fig:comparison_bmax_21cmfast}, we show the local variance of the noisy maps obtained in the optimistic and pessimistic case for the three 21CMFAST simulations: the three models can still be distinguished by the amplitude of $\sigloc$ when analysing noisy maps. We see that the smoothing due to the coarser angular resolution of the pessimistic case leads to a decrease in the signal. However, a lower angular resolution is also equivalent to a lower noise variance (see Eq.~\ref{eq:noise_variance}), such that the local variance of the noise does not exceed $0.02~\mathrm{mK}$, and the signal-to-noise ratio is around 50 during the bulk of reionisation. Therefore, we can recover the location of the maximum as well as the shape of the signal very well. Additionally, the ratio of the local variance maxima from one \texttt{rsage} model to another is well preserved: adding observational effects does not alter the ability of the local variance to differentiate between reionisation models.

Indeed, we fit a parabola $y(z)/\sigma_0 = \sigma_\mathrm{max}/\sigma_0 - (z-z_\mathrm{max})^2$, with $\sigma_0=1\mathrm{mK}$, to the local variance data points of our M9 simulation, on a redshift range $6.8 \leq z \leq 7.5$, for clean, noisy and corrected values. In this expression, $z_\mathrm{max}$ is the redshift when the maximum of the local variance is reached, and $\sigma_\mathrm{max}$ its amplitude. To estimate the corresponding uncertainties, we derive the standard deviation of the local variance by running 21CMFAST for identical physical and numerical parameters but $200$ different random seeds, which is equivalent to computing 200 different realisations of the simulation. We find that, in the optimistic case, the recovered amplitude is identical for all three data sets, giving $\sigma_\mathrm{max} = (0.82 \pm 0.24)~\mathrm{mK}$ at the $95\%$ confidence level. The location of the peak is slightly shifted towards larger redshifts for noisy data: we find $z_\mathrm{max}=7.1^{+0.3}_{-0.4} $ for both the optimistic and the pessimistic case, which is close to the value obtained for clean data ($z_\mathrm{max}=7.2^{+0.3}_{-0.4}$), and most importantly, within the size of a redshift bin (the redshift step between two snapshots is $\Delta z = 0.2$).

Accounting for the thermal noise is not sufficient to claim that our statistic will keep its characteristics and constraining power when analysing observed data, as we have not considered the impact of foreground avoidance or removal on our results. Nevertheless, \citet{HarkerZaroubi_2009} have found that one-point statistics are quite robust against the details of foreground fitting. In contrast, \citet{PetrovicOh_2011} have shown that foreground cleaning can significantly distort the one-point PDF by smoothing out its bi-modal structure\footnote{Ionised pixels form a Dirac peak centred on zero, whilst warm pixels are distributed more evenly.}. This will not be an issue for our local variance analysis, since, in contrast to the 3D pixel distribution, the distribution of mean values is smoothed by the averages along the frequency direction, and therefore not bi-modal. Additionally, foreground cleaning will reduce the contrast between neutral and ionised regions while maintaining the topology of the map, so that the distribution of means values and the variance within individual slices should be maintained. Finally, and similarly to our thermal noise analysis, it might be possible to remove the effects of foreground removal from the measured local variance, if we can characterise the statistical properties of foreground residuals sufficiently well. Analysing the contribution of the different $k$-modes to our local variance signal (for details see App.~\ref{app:contribution_modes}), we find small $k$-modes, for which foreground contamination is the largest, to contribute the most. We therefore discard the possibility of using foreground avoidance to derive the local variance from 21cm data. Instead foreground modelling and subtraction \citep{ChapmanAbdalla_2013,MertensGhosh_2018_GPR,HothiChapman_2021} or using machine-learning techniques to reconstruct the signal lost in the foreground wedge \citep{Gagnon-HartmanCui_2021}, should be preferred. We keep a thorough analysis of the impact of foreground removal on the local variance for future work. 

\subsection{Cross-correlations between slices}
\label{subsec:conclu_cross_correlations}

\begin{figure}
    \centering
    \includegraphics[height=6cm]{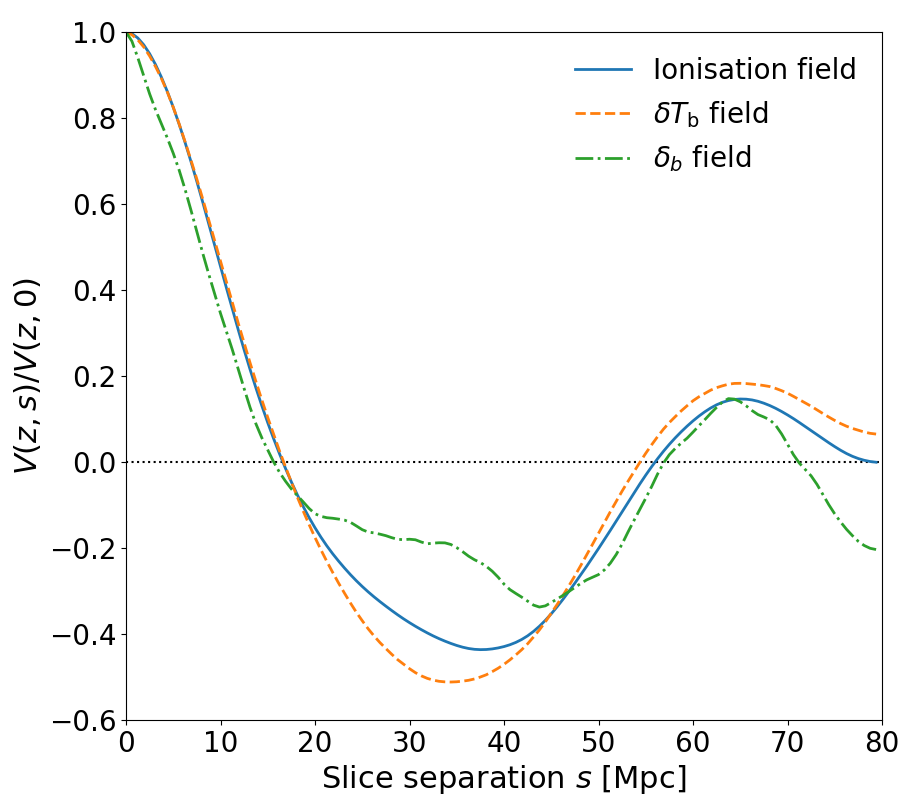}
    \caption{Cross-correlations between the average values of two slices cut through the M9 simulation and separated by a distance $s$, for the snapshot corresponding to $z=7.2$, normalised by the value at $s=0$. Results are presented for the ionisation field (solid blue line) and the 21cm brightness temperature field (dashed orange line).}
    \label{fig:corr_vs_dist_M9}
\end{figure}

In the previous sections, we have discussed the auto-correlations of slices within a simulation box. Another option is to analyse the cross-correlations between the average values of slices separated by a given distance $s$. This approach is similar to what has been done for the one-point PDF in \citet{BarkanaLoeb_2008, IchikawaBarkana_2010} and \citet{Gluscevic_Barkana_2010}. For the M9 simulation, we compute the following variance, to which we refer to as the cross variance:
\begin{equation}
    V(z,s)^2 = \frac{1}{N} \sum_{k=1}^N \mu(k,z)\, \mu(k+i_s,z) - \bar{x}_e^2,
\end{equation}
where $\mu(k,z)$ is the average of the $k^\mathrm{th}$ slice and $\mu(k+i_s,z)$ is the average of the $(k+i_s)^\mathrm{th}$ slice with $s=i\times \Delta x$ being the distance in $\mathrm{Mpc}$ separating them. This cross variance is equivalent to computing the 1D 2-point correlation function of the distribution of the means of slices. An example for the cross-correlation between slices at $z=7.2$ and $\bar{x}_e= 0.47$ is shown in Fig.~\ref{fig:corr_vs_dist_M9} for the density, ionisation and brightness temperature fields of the M9 simulation. We find that, for all fields, the cross correlation is maximal at $s=0$ , corresponding to auto-correlations, and decreases towards larger separations until it reaches negative values. A similar evolution was observed in \citet{munoz_2020_cosmic_variance}. %Because of the periodic boundary conditions of the simulation, a bump is visible when the separation becomes comparable to half the box size ($L/2=80~\mathrm{Mpc}$). 
We note that the values presented in the figure are normalised by $V(z,0)$. Raw values are of the order of $\sim 10^{-6}$ for the ionisation field and $0.1~\mathrm{mK}^2$ for the 21cm brightness temperature field. 

We compute the cross variance for all snapshots available in the M9 simulation. Interestingly, while the amplitude of $V(z,s)$ changes with redshift for the density field, its shape remains constant, and so does $V(z,s)/V(z,0)$. Initially, the cross variance of the $\delta T_b$ field has the same behaviour as the density field, until the box is about $10\%$ ionised. For the same redshifts, the scale at which $V(z,s)$ derived from the ionisation field becomes zero, $s_0(z)$, remains around $15~\mathrm{Mpc}$. At $\bar{x}_e \geq 40\%$, the $\delta T_b$ field mostly follows $x_\ion{H}{I}$  and $s_0$ starts increasing. The maximum of $s_0$ is reached at $\bar{x}_e = 90\%$. At this time, all ionised regions have percolated and the variance drops to zero. If we compare results between the three 21CMFAST simulations, we find their cross variances to be very similar, when they are measured at the same stage in the reionisation process. For this reason, it is not possible to use the cross variance to constrain reionisation physics. We keep a thorough investigation of these cross-correlations for future work.

The value of $s_0$ at $z=7.2$ is equivalent to about $1~\mathrm{MHz}$, which also corresponds to the order of magnitude of the coherency length of the cosmological 21cm signal during reionisation. While foregrounds are expected to have a much larger coherency length ($\sim 50-100~\mathrm{Mpc}$), noise has a zero coherency length \citep{SantosCooray_2005, MertensGhosh_2018_GPR}. These differences provide the basis for statistical 21cm signal separation via Gaussian Process Regression analysis \citep[GPR,][]{MertensGhosh_2018_GPR}, where the different components of the observed signal are modelled in order to remove foregrounds from the 21cm observations. Because their coherency length is much larger than the frequency bandwidth, the cosmological 21cm signal and foregrounds measured in two consecutive frequency channels (or, here, slices) should be almost identical and, when subtracting measurements in these two channels, only the uncorrelated thermal noise should remain. In \citet{PatilYatawatta_2016}, the authors use this technique to estimate the noise properties and remove noise from LOFAR data. Indeed, in contrast to foregrounds and cosmological signal, the thermal noise is found to be uncorrelated, even with a frequency separation as small as $0.2~\mathrm{MHz}$. Therefore, the difference between two Stokes I images in two consecutive frequency channels, after removing bright sources, will be dominated by thermal noise. Estimating the noise properties with this method leads to higher noise levels than when using Stokes V. The authors state that this excess noise is due to their calibration with an incomplete model. However, since this excess noise is uncorrelated between different observations, multiplying different observations will decrease the effective noise.

\section{Discussion \& Conclusion}
\label{sec:conclusions}

In this paper, we have presented a novel first-order statistics, the local variance $\sigloc$, which can be used to constrain the history and morphology of reionisation. 

The local variance corresponds to the variance of the distribution of means of slices taken along an axis in a simulation, or along the frequency direction of observations, if the channel width is sufficiently narrow.
At a fixed global ionisation level, the amplitude of the local variance of the ionisation field $\siglocion$ is a tracer of the size of ionised regions: it is higher for an ionisation field showing a few large ionised regions, that is when ionising sources are more scarce but more efficient at ionising. For a field made of many small ionised regions, as it arises when low-mass sources are the main drivers of reionisation for example, the local variance will be smaller. In future work, we will investigate in more details how the local variance can constrain the physical properties of early galaxies.
Additionally, the filling fraction for which the local variance reaches its maximum is mostly sensitive to the large-scale structure of the ionisation field during reionisation. It will be reached when both ionised and neutral regions are the largest, which occurs when approximately $60\%$ of the box is ionised. We have found that a more biased ionising emissivity distribution (that is fewer sources with higher ionising emissivities opposed to many sources with lower ionising emissivities) results in the maximum local variance being reached earlier in the reionisation process, that is at a lower ionisation fraction $\bar{x}_e$. 
Finally, when applying our novel statistics to the differential 21cm brightness temperature fields, the redshift-evolution of $\siglocbright$ exhibits a characteristic shape that traces the reionisation history of the sky patch observed. Before the onset of reionisation, $\siglocbright$ traces the correlations within the density field, but becomes sensitive to the ionisation morphology as a rising number of ionised regions emerge and grow.

We have shown that $\siglocbright$ is robust against thermal noise and angular resolution pollution when conducting $1000~$hrs observations with SKA1-\textit{Low}. For high angular resolution -- corresponding to a maximum baseline of $65~\mathrm{km}$, the smoothing due to limited angular resolution has no impact on the local variance but the thermal noise adds amplitude and fluctuations to the signal. We have found that if the statistical properties of the noise are sufficiently known, the thermal noise contribution can be removed from the measured signal, and only the thermal noise fluctuations are conserved. For lower angular resolution -- corresponding to a maximum baseline of $2~\mathrm{km}$, the smoothing leads to a decrease in the variance of the field and the noise level. In both cases, we can recover the maximum of the local variance as well as differentiate between the different reionisation models investigated in this work from noisy 21cm maps. We expect this result to hold even after foreground removal. Indeed, $\sigloc$ changes only weakly when the one-point PDF is altered by foreground removal \citep{PetrovicOh_2011}. A detailed analysis of the impact of foreground removal on the local variance will be the focus of future work.
In conclusion, local variance data will enable the recovery the reionisation history in a model-free fashion, as well as the constraint of astrophysical parameters related to the physical properties of early galaxies. Applying our local variance statistics to measurements would consist of obtaining the reionisation midpoint of a statistical sample of small fields of view, which are either obtained by dividing a larger observational window into smaller areas or by performing a series of different observations, and combining these `local' reionisation midpoints to estimate a global reionisation midpoint. In this work, for computational reasons, individual observations have been represented by the different slices in a single coeval cube. %Which of these observational strategies would be optimal will be investigated in future work.

We note that other sample shapes could have been considered instead of slices. For example, \citet{GiriD'Aloisio_2019} consider the power spectrum measured in sub-cubes of a simulation. Since observational data cubes will consist of slices at different (but gradually increasing) frequencies, our findings can be easier translated to the results from observations. Alternatively, \citet{BittnerLoeb_2011} consider the distribution of means measured across beams drawn along the frequency direction, and one could also consider the mean of sub-cubes in the entire coeval simulation box. Preliminary calculations have shown that the local variances of both shapes suffer from the same drawbacks, that is their strong dependency on the simulation size and resolution, but yield very similar results.

As the integral of the power spectrum, the local variance inherently includes the same information about the underlying reionisation field. However, it will be less affected by observational limitations. For example, because it is measured at a given frequency, we can avoid frequency channels contaminated by radio frequency interference (RFI). In addition, extracting patches of sky smaller than the total field of view of the 21cm observations can eliminate survey boundary effects coming from tapering. In App.~\ref{app:comparison_PS}, we find the local variance to be more robust to thermal noise than the power spectrum, except on scales $k < 0.5~\mathrm{Mpc}^{-1}$, where the cosmological signal is expected to be swamped by foregrounds.

 Depending on the reionisation scenario, the 21cm local variance can reach values as high as $1~\mathrm{mK}$. However, this value will naturally decrease as larger field of views or higher frequency resolutions are considered. While in this work, large values of the sample variance are desired, it has previously been considered an issue, as it represents an obstacle for a precise estimate of the 21cm global signal or power spectrum and, in turn, of astrophysical parameters. \citet{munoz_2020_cosmic_variance} proposed a way of quickly estimating the sample variance when measuring the 21cm global signal as a function of its derivative with respect to redshift. Considering a simulation as big as $L=1.8~\mathrm{Gpc}$, \citet{munoz_2020_cosmic_variance} find a maximum sample variance of $\sigma_{21} \sim 0.6~\mathrm{mK} $ at $z=16.8$. In parallel, \citet{kaur_2020} estimated that a simulation needs to be at least $200-300~\mathrm{Mpc}$ wide to obtain a bias-free 21cm power spectrum on scales of $-1.2 < \log k / \mathrm{Mpc}^{-1} < 0$. In contrast to other estimators, the local variance uses sample variance, often considered an observational bias, to our benefit. Looking for optimal observational strategies allowing a maximum amplitude of the local variance, while minimising its error bars, in order to offer reliable constraints, will be the focus of future work.

\begin{acknowledgements}

The authors thank the referee for useful comments on this manuscript, which helped improve its overall quality. They also thank Catherine A. Watkinson, Ian Hothi, Adrian C. Liu and Jordan Mirocha for their input on a draft version of this paper; as well as Jacob Seiler for providing the \texttt{rsage} simulations. 
AG and JP acknowledge financial support from the European Research Council under ERC grant number StG-638743 ("FIRSTDAWN"). AG's work was additionally supported by the McGill Astrophysics Fellowship funded by the Trottier Chair in Astrophysics, as well as the Canadian Institute for Advanced Research (CIFAR) Azrieli Global Scholars program. AH acknowledges support from the European Research Council's starting grant ERC StG-717001 ("DELPHI").  The idea of this work was developed thanks to visits between the authors of this paper, partly funded by the Leids Kerkhoven-Bosscha Fonds (LKBF). \\

This research made use of \texttt{astropy}, a community-developed core Python package for astronomy \citep{astropy,astropy2}; \texttt{matplotlib}, a Python library for publication quality graphics \citep{hunter_2007}; and of \texttt{scipy}, a Python-based ecosystem of open-source software for mathematics, science, and engineering \citep{scipy} -- including \texttt{numpy} \citep{numpy}.
\end{acknowledgements}

\bibliographystyle{aa} 
\bibliography{biblio} 

\appendix

\section{Local variance of the $\delta T_\mathrm{b}$ field}
\label{app:dTb_var_calc}

\begin{figure}
    \centering
    \includegraphics[height=6cm]{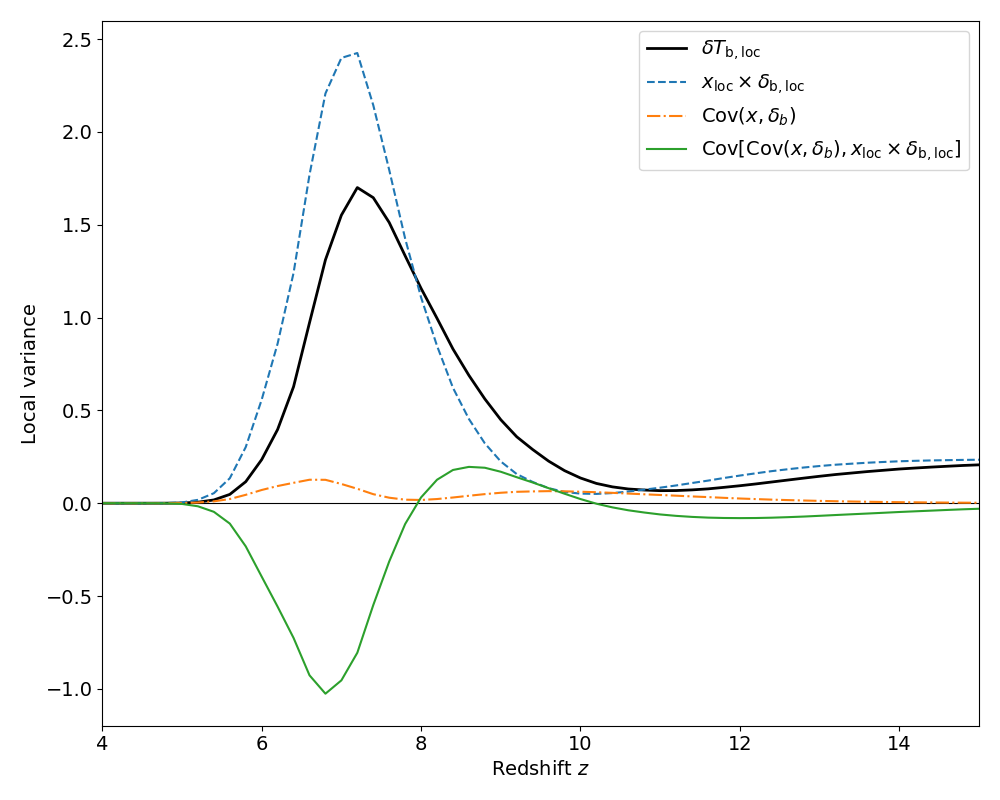}
    \caption{Contributions to the local variance of the 21cm brightness temperature field in a 21CMFAST simulation.}
    \label{fig:app_contributions}
\end{figure}

In the following we consider the 21cm brightness temperature field to be the direct product of the neutral hydrogen \ion{H}{I} and the overdensity $\delta_\mathrm{b} = \rho_\mathrm{b}/\bar{\rho}_\mathrm{b} -1$ fields,
\begin{equation}
    \delta T_\mathrm{b}(\bm{x}) = x_\ion{H}{I}(\bm{x}) \times \left[1+\delta_\mathrm{b}(\bm{x})\right].
\end{equation}
If we consider the $k$th slice along the frequency direction in the simulation, we yield from the previous equation
\begin{equation}
    \label{eq:deltaTbLoc_freq}
    \delta T_\mathrm{b,loc}(k) = E[\delta T_\mathrm{b}(k)] = E \left[ x_\ion{H}{I}(k) \times (1+\delta_\mathrm{b}(k))  \right],
\end{equation}
where $E$ is the expectation value, $\delta T_\mathrm{b,loc}(k)$ is the mean of the 21cm field of the slice, $x_\ion{H}{I}(k)$ its 2D $\ion{H}{I}$ field and $\delta_\mathrm{b}(k)$ its 2D overdensity field. Since these two fields are correlated, equation \ref{eq:deltaTbLoc_freq} can be reformulated as
\begin{equation}
\begin{aligned}
    \delta T_\mathrm{b,loc}(k) = \bar{x}_\ion{H}{I}&(k) \times  (1+\bar{\delta}_\mathrm{b}(k)) \\ & + \mathrm{Cov}\left[ x_\ion{H}{I}(k) \times (1+\delta_\mathrm{b}(k))  \right].
\end{aligned}
\end{equation}
Then the local variance is the variance of the distribution of $\delta T_\mathrm{b,loc}$ values. If we write $X=\bar{x}_\ion{H}{I}(k) \times  (1+\bar{\delta}_\mathrm{b}(k))$ and $Y=\mathrm{Cov}\left[ x_\ion{H}{I}(k) \times (1+\delta_\mathrm{b}(k))  \right]$ then
\begin{equation}
\begin{aligned}
   \sigloc^2 & =  \mathrm{Var}(\delta T_\mathrm{b,loc}) \\ 
    & = 2~\mathrm{Cov}(X,Y) + \mathrm{Var}X + \mathrm{Var}Y\\
    & = \mathrm{Cov}[\bar{x}_\ion{H}{I}^2, (1+\bar{\delta}_\mathrm{b})^2] + \left( \sigma_{\mathrm{loc},\ion{H}{I}}^2 + \bar{x}_\ion{H}{I}^2 \right) \left( \sigma_{\mathrm{loc},\delta_b}^2 + 1 \right)\\ & - \left[ \mathrm{Cov}[\bar{x}_\ion{H}{I}, (1+\bar{\delta}_\mathrm{b})] + \bar{x}_\ion{H}{I} \right]^2.
\end{aligned}
\end{equation}
These different elements are represented in Fig.~\ref{fig:app_contributions}.
 We see that the final shape of the local variance is mostly made of the variance of $X$, the product of the two local fields, and of the covariance of $X$ and $Y$, which are both complicated objects difficult to interpret. However, as shown in Fig.~\ref{fig:varloc_tb_contributions}, considering the local variances of $x_\ion{H}{I}$ and $1+\delta_\mathrm{b}$, as well as their covariance, is sufficient to understand the redshift-evolution of $\siglocbright$.

\section{Tests on toy models}
\label{app:toy_model}

\begin{figure}
    \centering
    \includegraphics[height=6cm]{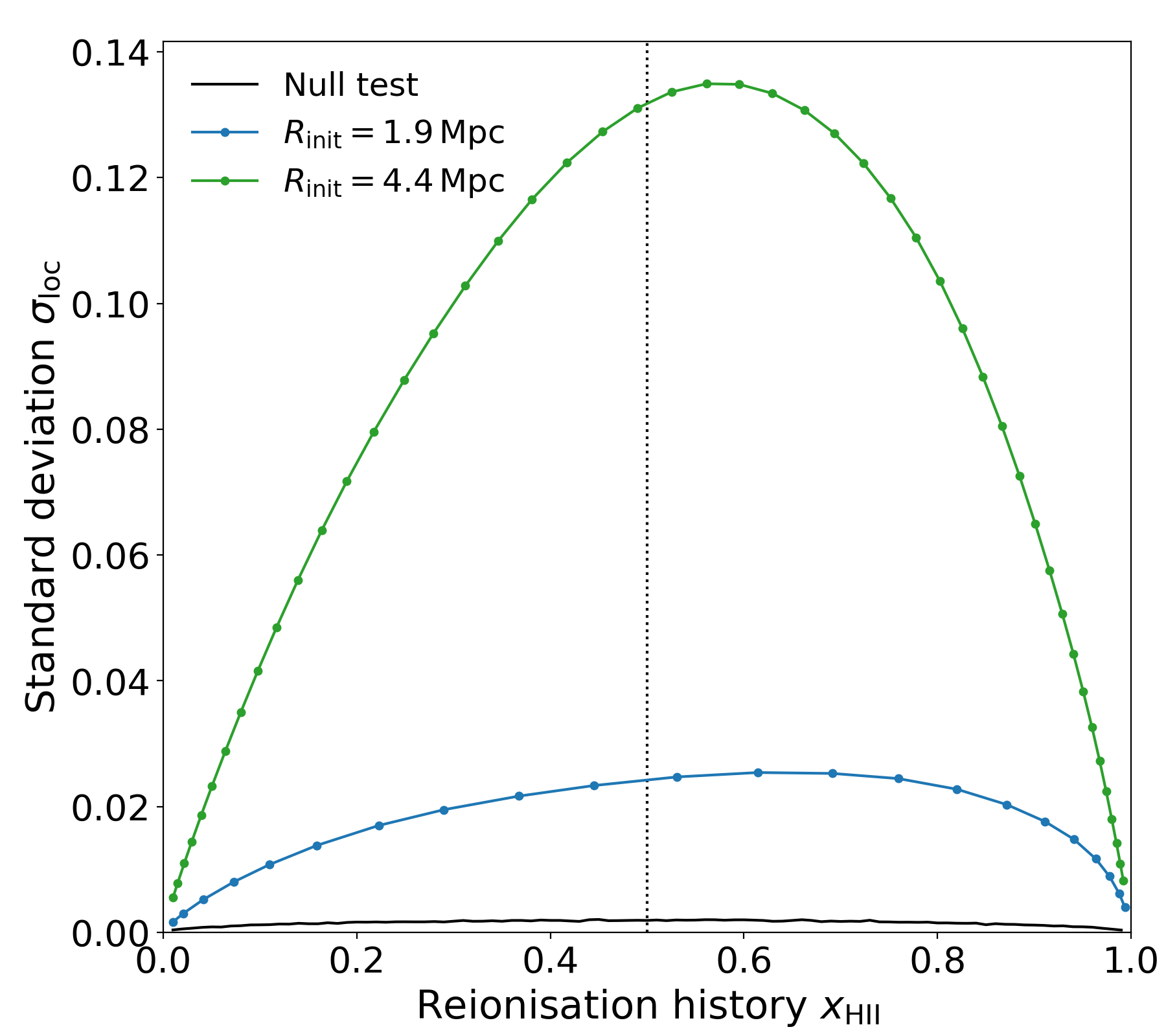}
    \caption{Local variance $\sigloc$ for toy models with different starting radius.}
    \label{fig:app_toy_models}
\end{figure}

We generate toy models with dimensions equal to those of the \texttt{rsage} simulations, that is a comoving box length of $L=160~\mathrm{Mpc}$ and 256 grid cells on each side. The initial field is a 3D neutral box filled with enough bubbles of radius $R_\mathrm{init}$ to reach a filling fraction of $\bar{x}_e = 0.01$, which are then artificially grown by 1 cell in radius until a filling fraction of $100\%$ is reached. We compute the local variance for each of the resulting boxes. Results can be seen in Fig.~\ref{fig:app_toy_models} in comparison to a control test where ionised pixels are randomly distributed. We evolve the toy model boxes with different initial radii: a large $R_\mathrm{init}$ will be equivalent to a higher mass threshold for ionising sources, that is an increasing $R_\mathrm{init}$ will be equivalent to an increasing $M_\mathrm{turn}$ in 21cmFAST or transitioning from \texttt{rsage fej} via \texttt{rsage const} to \texttt{rsage SFR}. The difference to 21CMFAST and \texttt{rsage} is that ionised bubbles are randomly located, so the ionising sources are not clustered. Additionally, there are no new ionised regions throughout the process, since all the bubbles are initialised in the first field. Despite these differences, the shape of $\siglocion(x_e)$ remains unchanged, and the maximum is still reached for a filling fraction of about $0.60$: $\bar{x}_e=0.61$ for the field with small bubbles, and $\bar{x}_e=0.56$ for the field with large bubbles. Similarly to what has been seen from the 21CMFAST simulations, the simulation with the largest bubbles reaches on average its maximum local variance at smaller filling fractions.

\section{Dependence on box size and other limitations}
\label{subsec:discussion_box_size}

\begin{figure}
    \centering
    \includegraphics[height=6cm]{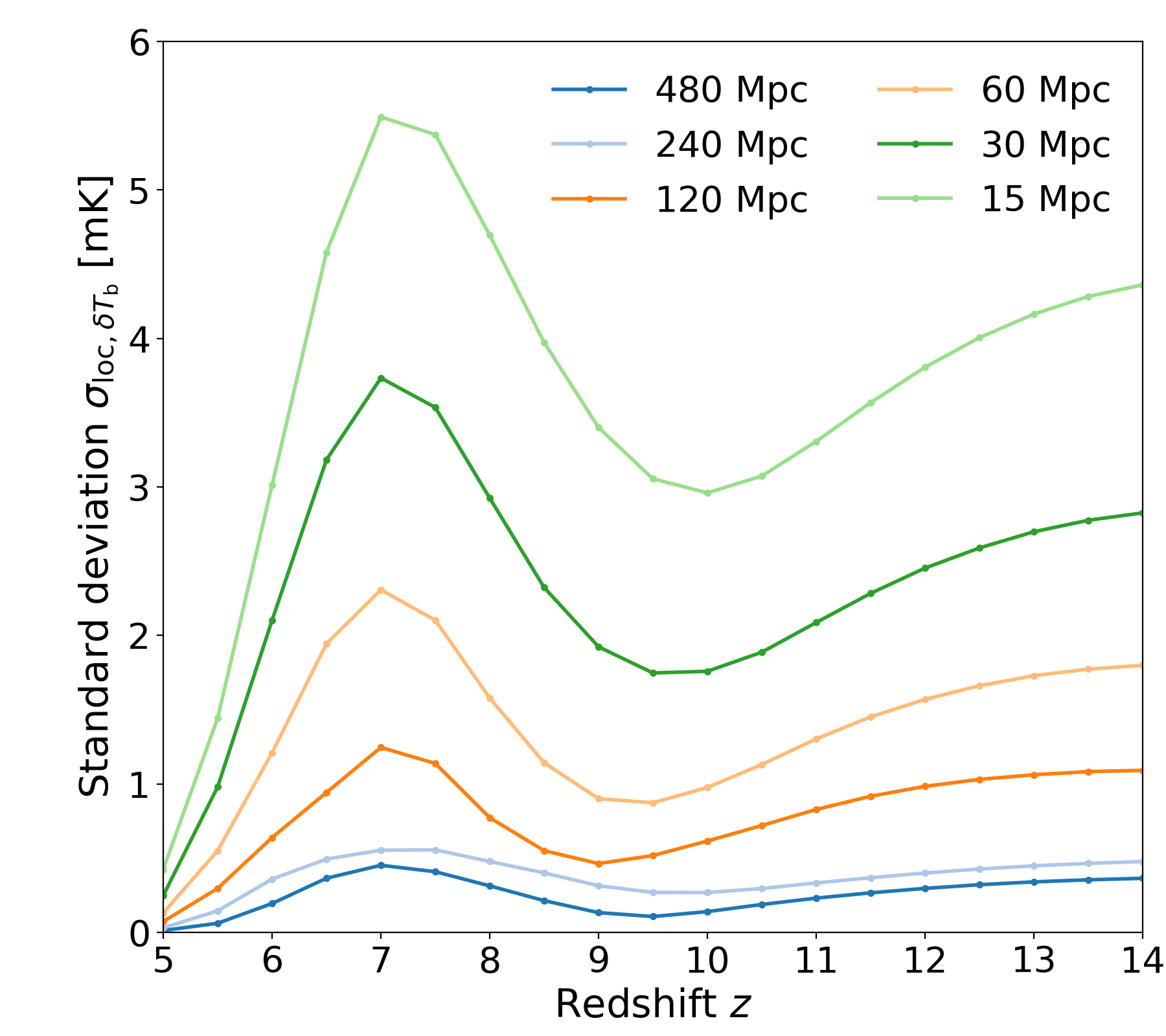}
    \caption{Evolution of the standard deviation of the $\xloc$ distributions for  M9 simulations of decreasing side length but constant resolution.}
    \label{fig:box_size_M9}
\end{figure}

Because the information contained in $\sigloc$ is purely related to the sample variance, $\sigloc$ will depend on the size of the box considered: the larger the box, the smaller the variance. In order to compare with the fields-of-views anticipated for upcoming 21cm experiments, we ask what is the limiting size that allows us to differentiate between reionisation models with $\sigloc$. To do so, we generate a 21CMFAST simulation box for $\Mturn=10^9~M_\odot$, side length $L=480~\mathrm{Mpc}$ and cell size $\Delta x = 0.625~\mathrm{Mpc}$, same as before. We divide this large simulation into sub-cubes of decreasing size, until a side length of $L=15~\mathrm{Mpc}$ is reached (All sub-cubes have identical reionisation histories.). We compute the local variance obtained from the 21cm brightness temperature fields of all sub-cubes and compare their values in Fig.~\ref{fig:box_size_M9}. The maximum signal is reached for the smallest box size, $L=15~\mathrm{Mpc}$, and reaches values as high as $5.5~\mathrm{mK}$, which is 10 times higher than the fiducial boxes we analysed in section \ref{app:toy_model}. For all box sizes, we can locate a maximum signal at the same redshift, corresponding to $61\%$ of global ionisation level. It decreases drastically with box size, according to 
\begin{equation}
    % \mathrm{amp}\ \sigloc_{,\delta T_\mathrm{b}}/\mathrm{mK} \simeq - 4.1 \log (L/\mathrm{Mpc}) + 10.0,
    \sigloc_{,\delta T_\mathrm{b}} \sim 1.3~\mathrm{mK}\times \left( \frac{L}{100~\mathrm{Mpc}}\right)^{-0.8}.
\end{equation}
SKA1-\textit{Low} has a field of view of $327~\mathrm{arcmin}$ at nominal frequency of $110~\mathrm{MHz}$, which corresponds to $L=760, 840, 900, 940$, and $970~\mathrm{Mpc}$ at $z= 5,7,9,11$ and $13$, respectively. 
For such wide observational windows, the sample variance is weak: at $z=6$, towards the end of reionisation, it will be about $\sim 0.25~\mathrm{mK}$. 
This relation is only valid at a time when about $60\%$ of the IGM is ionised, which will correspond to different redshifts depending on the reionisation scenario. However, it will give the maximum amplitude of $\sigloc$ and is therefore a useful choice, if we want to estimate the errors on 21cm global signal measurements. Interestingly, it is about 10 times smaller than the one found by \citet{munoz_2020_cosmic_variance} at $z=16.3$, which confirms how dependent the amplitude of the local variance is on the physics of reionisation and on the reionisation stage it is computed at. %their resolution is $3~\mathrm{Mpc}$

Additionally, we expect $\sigloc$ to depend on the resolution of the simulation considered -- or on the angular resolution of the telescope used. Indeed, \citet{BanetBarkana_2020} already pointed out the impact of instrument resolution and smoothing on the one-point and differential PDF. The anticipated angular resolution of SKA1 at a nominal frequency of $110~\mathrm{MHz}$ is expected to be $11~\mathrm{arcsec}$ \citep{BraunBonaldi_2019}, that is between $0.45$ and $0.55~\mathrm{Mpc}$ on the redshift range $ 5 \leq z \leq 13$. Running 21CMFAST for $\Mturn=10^9~M_\odot$ and a fixed number of cells but an increasing resolution, from $\Delta x = 0.5~\mathrm{Mpc}$ to $3~\mathrm{Mpc}$, we find that an improved resolution enhances the local variance as structures are better resolved and the number of partially ionised regions decreases. 

Finally, computing $\sigloc$ requires to have a sufficient number of slices available: it is obvious from Fig.~\ref{fig:rsage_const_slices} that without a sufficient number of slices, the $\xloc$ distributions will be noisy and their variance $\sigloc$ will not be reliable. For the M9 simulation, we find that 128 slices, so half of the box, will still provide satisfying results, while lower sample sizes make $\sigloc$ unusable. However, thanks to the very high frequency resolution of SKA, anticipated to be $5.4~\mathrm{kHz}$ at $110~\mathrm{MHz}$ \citep{BraunBonaldi_2019}, we expect to observe a sufficient number of images at sufficiently close redshifts for $\sigloc$ to give interesting results. This will be the focus of future work.

\section{Mode contribution to the local variance}
\label{app:contribution_modes}

\begin{figure}
    \centering
    \includegraphics[width=.9\linewidth]{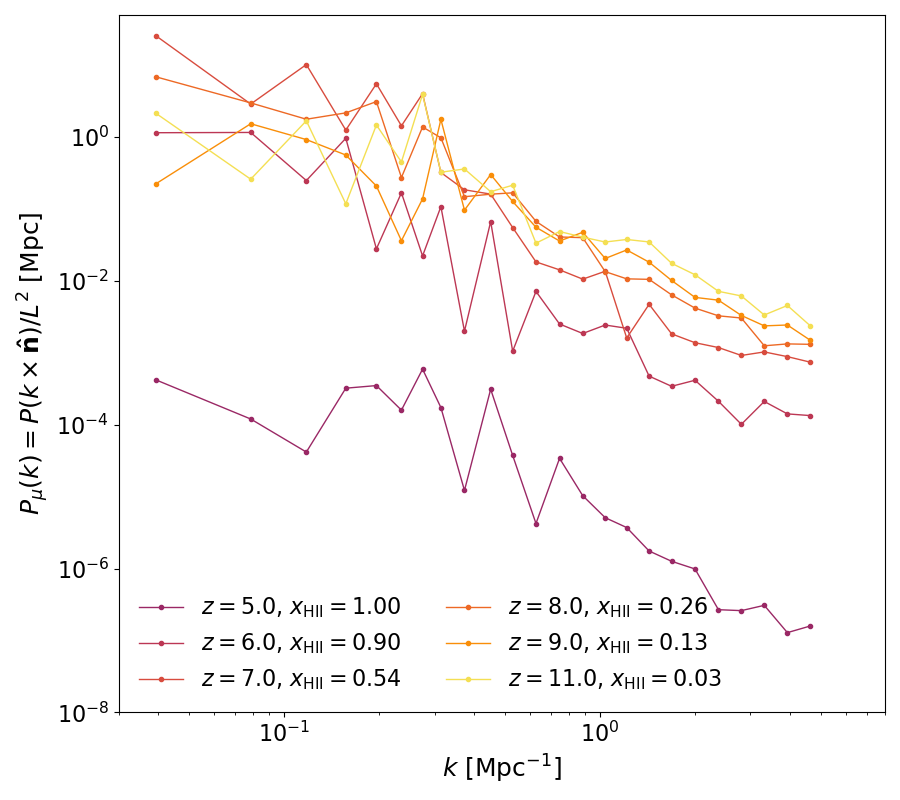}
    \caption{Evolution of the power spectrum of the 1D distribution of means along the redshift direction $P_\mu(k)$ of the M9 simulation, illustrating the contribution of Fourier modes to the local variance.}
    \label{fig:app_contributions_modes}
\end{figure}

To estimate the impact of foregrounds on the local variance, we analyse the contribution of different $k$-modes to the $\sigloc(z)$ signal. According to Eq.~\ref{eq:sigloc_vs_PS}, the local variance is the integral over the power spectrum of the 1D distribution of the mean values along the line of sight and/or redshift direction $P_\mu(k)$. Hence, $P_\mu(k)$ is a direct measure of the contribution of each $k$-mode to the local variance: for the M9 simulation, we show $P_\mu(k)$ at $z=5-11$ in Fig.~\ref{fig:app_contributions_modes}.

First, and as expected, the evolution of the global amplitude of $P_\mu(k)$ recovers the redshift-evolution of the local variance and, in particular, its maximum around $z\sim 7$. Additionally, we see that the low-$k$ modes are the ones contributing the most to the local variance signal as $P_\mu(k)$ decreases with increasing $k$ values following $P_\mu(k) \propto k^{-2}$ approximately at all redshifts. This confirms our findings in Sec.~\ref{subsec:comparison_sigloc}, that the local variance, and in particular the location of its maximum, is sensitive to the ionisation morphology on large scales. Furthermore, because foreground corruption is larger for small $k$-modes, this result shows that foreground avoidance is likely to diminish the reionisation signal in the local variance of the 21cm signal. Other possibilities should be considered, such as foreground removal \citep{ChapmanAbdalla_2013,HothiChapman_2021} or machine-learning techniques to reconstruct the wedge \citep{Gagnon-HartmanCui_2021}.

\section{Performance comparison with the power spectrum}
\label{app:comparison_PS}

As an integral of the power spectrum, the variance -- local or not, inherently encompasses the same information. However, as we explain in this Section, we expect the local variance o be less affected by observational limitations than the power spectrum.

In Sec.~\ref{sec:sims} we have mentioned that, if this work was based on the analysis of coeval cubes for computational reasons, this approach is not directly transferable to 21cm observations since different slices along the observed light cone correspond to different redshifts. Instead, we would consider a wide field-of-view at a given frequency (or redshift), and divide it into sub-patches -- an approach already imagined in \citet{GiriD'Aloisio_2019}. Comparing the means of these sub-patches is then equivalent to comparing the means of slices of a coeval cube. Therefore, one is able to pick sub-patches in a way that avoids survey boundary effects, which are a common problem of power spectrum estimations. Additionally, since the signal is measured for a given frequency bin, one can conveniently choose what frequency is considered, and in particular avoid frequency bands dominated by, for example, radio-frequency interference (RFI). 

\begin{figure}
    \includegraphics[width=.49\textwidth]{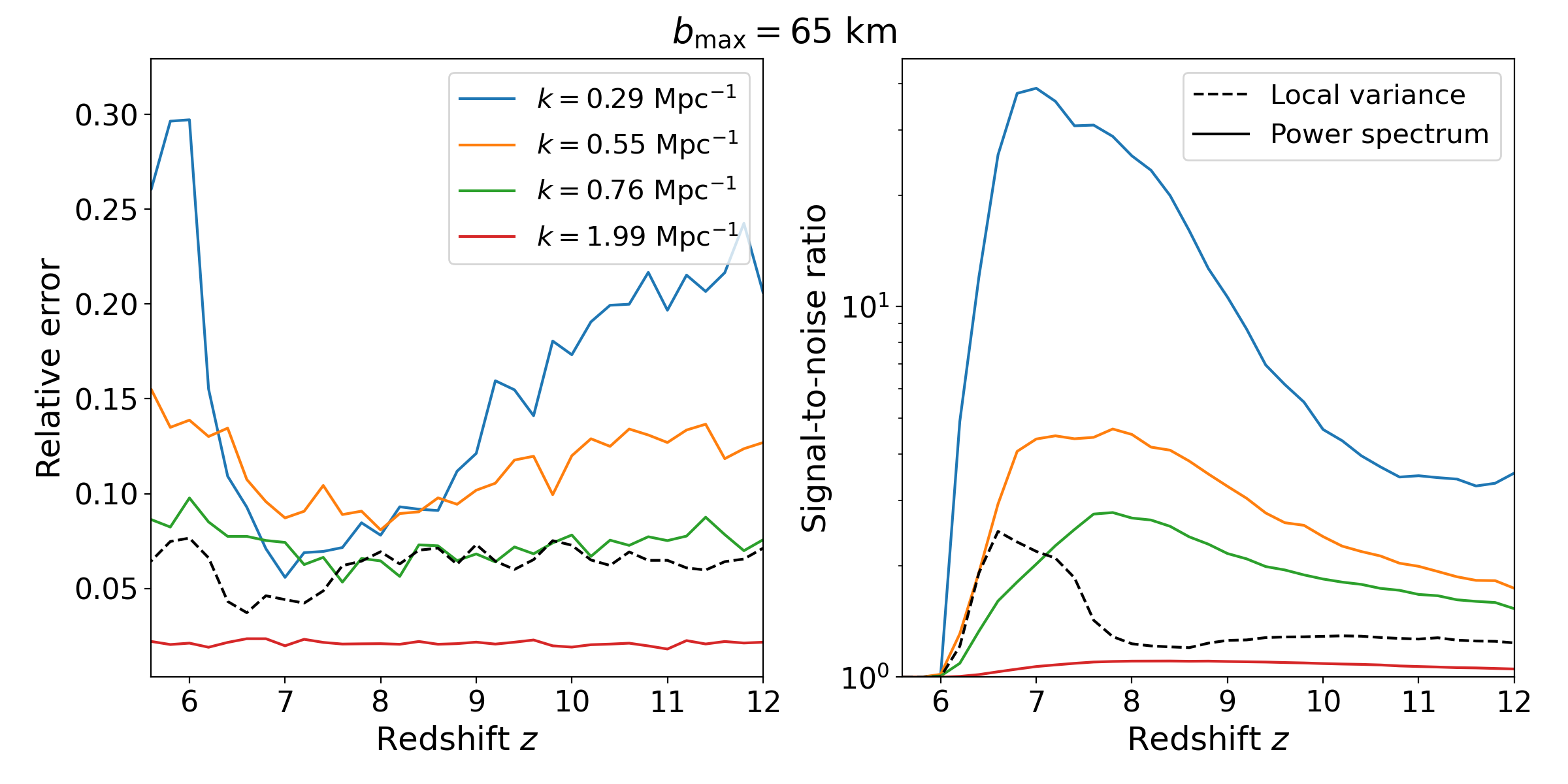}\\
    \includegraphics[width=.49\textwidth]{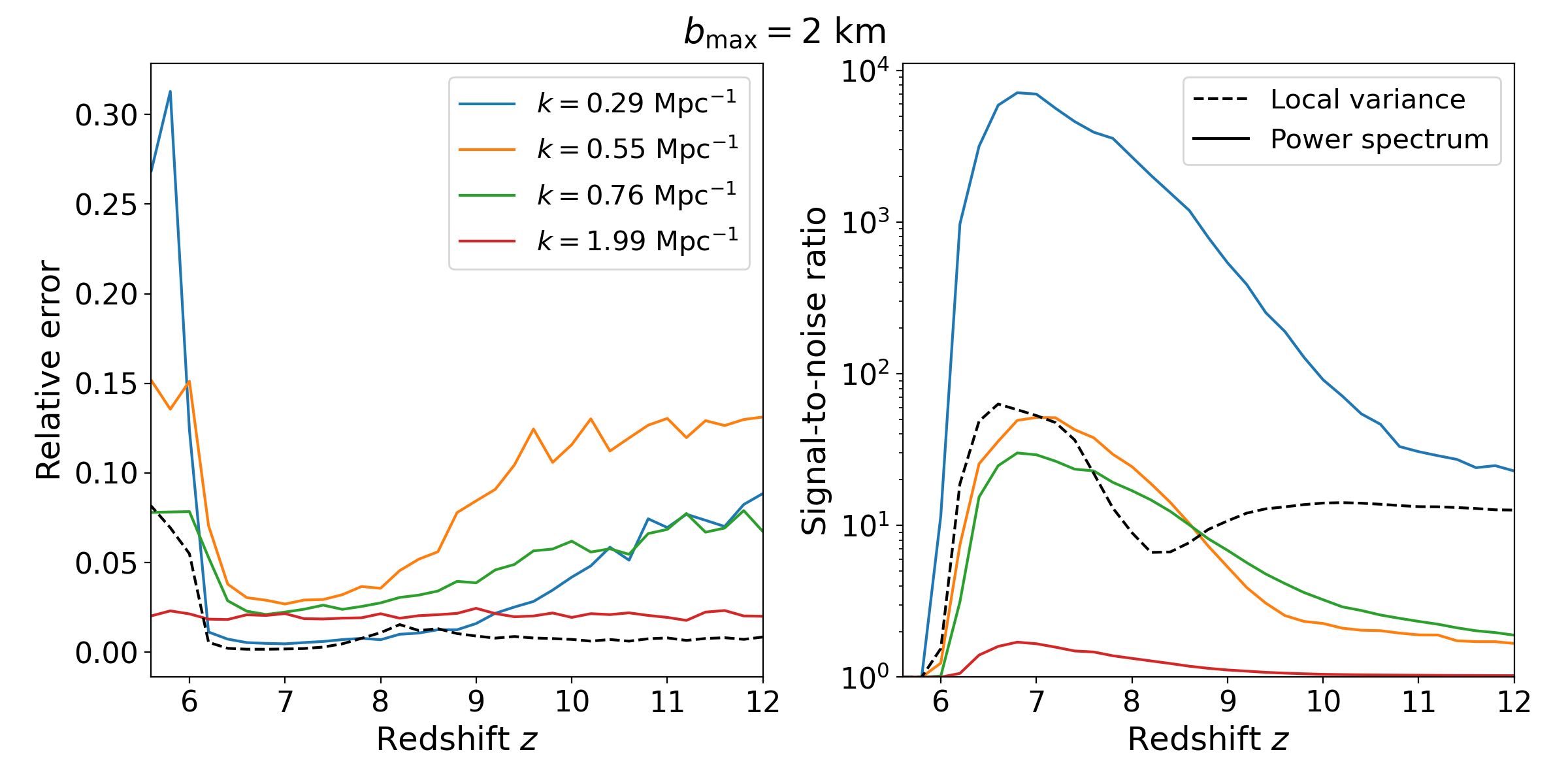}
    \caption{Relative error bars (left panel) and signal-to-noise ratio (right panel) obtained from measurements of the local variance and the 21cm power spectrum with different realisations of the noise added to a 21CMFAST simulation. Two observational cases are considered: a maximum baseline of $b_\mathrm{max}=65~\mathrm{km}$ (upper panel), and one of $b_\mathrm{max}=2~\mathrm{km}$ (lower panel).}
    \label{fig:compare_SNR_errors}
\end{figure}

We now perform a comparison of the robustness of the power spectrum and of the local variance to thermal noise. For every snapshot available from the M9 simulation, we generate 100 realisations of thermal noise fields and add them to the 21cm brightness temperature field that is smoothed according to a maximum baseline of $b_\mathrm{max}=2~\mathrm{km}$ or $65~\mathrm{km}$. We choose a 21CMFAST simulation of box size of $50~\mathrm{Mpc}$ for 100 pixels per side and set $\Mturn=10^9~M_\odot$. We compute the three-dimensional power spectrum and the local variance of each of these 100 boxes, and take their standard deviation as the error due to noise for each estimator. The resulting relative error bars are shown in the left panels of Fig.~\ref{fig:compare_SNR_errors}. We see that, thanks to $\sigloc$ being the integral over all $k$-modes, the relative error of the $\sigloc$ values remains constant over the redshift range, with values of $\sim6\%$ for the $b_\mathrm{max}=65~\mathrm{km}$ case. For $b_\mathrm{max}=2~\mathrm{km}$, results are even better with relative errors lower than $1\%$ for $z\geq 6$ and lower than the relative error obtained for the power spectrum on all scales. This is not surprising, since Eq.~\ref{eq:noise_variance} shows that the variance of the noise is inversely proportional to the angular resolution of the telescope and therefore smaller for our pessimistic case.

Additionally, we compute the power spectrum and local variance of the 100 realisations of thermal noise alone and derive the signal-to-noise ratio (SNR) of the two estimators as $\langle P_\mathrm{smoothed}(k) / P_\mathrm{noise}(k)\rangle$ and $\langle\sigloc_\mathrm{, smoothed} / \sigloc_\mathrm{, noise}\rangle$, respectively (see right panel in Fig.~\ref{fig:compare_SNR_errors}). For the reasons mentioned above, the SNR values of the local variance for the $b_\mathrm{max}=2~\mathrm{km}$ case exceed those of the pessimistic baseline case. They are maximum at the redshifts where the local variance reaches its maximum ($z\simeq6.5-7.5$), which is the range of greatest interest to us. For the pessimistic case, the SNR of the local variance is equal to $\sim 60$ at $z\simeq6.5-7.5$. We note that the SNR of the local variance is only smaller than the SNR of the power spectrum for small $k$-modes, $k < 0.5~\mathrm{Mpc}^{-1}$, but these modes are expected to be swamped by foregrounds. 

In summary, we find that on average our estimator is more robust to thermal noise than the power spectrum, especially on scales $k > 0.5~\mathrm{Mpc}^{-1}$ that are key for deriving constraints on the reionisation morphology from observations.

\end{document}